\documentclass[%
 reprint,
superscriptaddress,
preprintnumbers,
nofootinbib,
 amsmath,amssymb,
 aps,
]{revtex4-1}

\usepackage[colorlinks]{hyperref}
\usepackage{graphicx}
\usepackage{dcolumn}
\usepackage{bm}
\usepackage{color}
\usepackage{placeins}
\usepackage{slashed}
\usepackage{multirow}
\usepackage{bbold}

\usepackage{soul}

\definecolor{red}{rgb}{0.9, 0,0}
\definecolor{cerulean}{rgb}{0., 0.62,0.9}
\definecolor{navy}{rgb}{0.05, 0.05,0.8}

\renewcommand{\eqref}[1]{Eq.~\ref{#1}}

\newcommand{\eoff}{\epsilon_{\mathrm{off}}}
\newcommand{\etrig}{\epsilon_{\mathrm{trig}}}
\newcommand{\wacc}{w_{\mathrm{acc}}}
\newcommand{\met}{E\!\!\!/_T}

\begin{document}

\title{Scouting for dark showers at CMS and LHCb}

\author{Susan Born}
\affiliation{Department of Physics, University of Illinois  Urbana-Champaign, Urbana, IL 61801}
\author{Rohith Karur}
\affiliation{Theoretical Physics Group, Lawrence Berkeley National Laboratory, Berkeley, CA 94720, USA}
\affiliation{Berkeley Center for Theoretical Physics, Department of Physics, University of California, Berkeley, CA 94720, USA}
\author{Simon Knapen}
\affiliation{Theoretical Physics Group, Lawrence Berkeley National Laboratory, Berkeley, CA 94720, USA}
\affiliation{Berkeley Center for Theoretical Physics, Department of Physics, University of California, Berkeley, CA 94720, USA}
\author{Jessie Shelton}
\affiliation{Department of Physics, University of Illinois  Urbana-Champaign, Urbana, IL 61801}
\affiliation{Illinois Center for Advanced Studies of the Universe, University of Illinois Urbana-Champaign, Urbana, IL 61801} 
\affiliation{Center for Theoretical Physics, Department of Physics, Massachusetts Institute of Technology,
Cambridge, MA 02139}

\preprint{MIT-CTP/5537}

\makeatletter
\def\l@subsection#1#2{}
\def\l@subsubsection#1#2{}
\makeatother

\date{\today}

\begin{abstract}

We assess the capabilities of the CMS and LHCb searches for low-$p_T$ displaced dimuon pairs to discover hidden valley models. To do so, we develop a new benchmark model featuring a light dark photon with dark flavor-violating couplings, which realizes a range of dimuon vertex topologies.  We show that the data scouting techniques used in these searches provide unique sensitivity and we make some additional suggestions to further extend the scope of future experimental searches.

\end{abstract}

\maketitle

\tableofcontents

\section{Introduction}

In its second decade of operation the Large Hadron Collider (LHC) is now transitioning to its high-intensity phase, which will deliver substantial new sensitivity for precision Standard Model (SM) measurements as well as new opportunities to search for very rare beyond-the-SM phenomena in the energy range from the GeV scale up to a few hundreds of GeV.  In relation to the search for new physics, it has long been appreciated that trigger design can and should play a dynamical role in developing analyses, as opposed to a static obstruction when searching for soft signatures. All experiments therefore strive to perform as much event reconstruction and analysis as possible at the trigger level, as this allows for more informed decisions as to which events to commit to tape. 

Because more and more of the event reconstruction can be performed in real time, ATLAS, CMS and LHCb have also been able to record partial or reduced versions of the events reconstructed by their trigger-level algorithms. This allows for a much higher output rate, and thus greatly reduced trigger thresholds. This strategy is referred to as the Turbo Stream \cite{Aaij:2016rxn,Aaij:2019uij}, Data Scouting \cite{CMS:2016ltu} or Trigger-Level Analysis \cite{ATLAS:2018qto} by LHCb, CMS and ATLAS respectively. At LHCb in particular, this online-analysis concept is a critical component of their ongoing and future data taking strategies, with an $\mathcal{O}(1)$ fraction of all events already being recorded through the Turbo Stream \cite{Aaij:2019uij}. This has enabled LHCb to measure (for example) certain exclusive charm decays and allowed them to perform searches for new, low mass dimuon resonances \cite{Aaij:2017rft,Aaij:2019bvg,Aaij:2020ikh}. Fully online strategies have also been deployed to great effect in searches at ATLAS and CMS, in particular for low-mass dijet \cite{CMS:2016ltu,ATLAS:2018qto} and (displaced) dimuon resonances  \cite{PhysRevLett.124.131802,CMS:2021sch,CMS-PAS-EXO-21-005}. A scouting search for low mass diphoton resonances would furthermore be very well motivated \cite{Knapen:2021elo}, though may be technically challenging. Broadly speaking, data scouting strategies will become even more powerful tools during the HL-LHC phase, provided that the existing capabilities can be maintained and, optimally, expanded, in the context of the higher instantaneous luminosity planned for future operation. 

Data scouting is moreover of great interest from a theoretical point of view.  The emerging appreciation that new physics may be relatively light and coupled only feebly to the SM
(see e.g.~the reviews~\cite{Alexander:2016aln,Lanfranchi:2020crw})
places a renewed emphasis on searches for new phenomena in relatively low-energy final states.  Data scouting techniques are very well suited to  making these challenging, high-rate kinematic regimes accessible. Moreover, searches relying on data scouting are \emph{inherently inclusive}, insofar as (e.g.) a dimuon scouting trigger will record information about all dimuon pairs passing its selection criteria, independent of the properties of the rest of the event.  This property makes data scouting searches particularly suitable to capture a wide range of unexpected phenomena.  These two features are particularly important in the current and upcoming phases of LHC operations, 
as theory priors are nowadays much weaker than those during the initial phase of the LHC.

Hidden valley (HV) models are a well-motivated and physically rich class of dark sector models \cite{Strassler:2006im}, whose defining feature is the existence of interactions within their dark sector that generate a high-multiplicity final state. This is typically due to the presence of a confining gauge group in the dark sector that gives rise to  dark showering and hadronization, in broad analogy to the QCD sector in the Standard Model (SM). 
Hidden valleys specifically have been developed as components of models seeking to address the origin of dark matter \cite{Hur:2007uz,Kribs:2009fy,Beauchesne:2018myj,Francis:2018xjd,Bernreuther:2019pfb}, the matter-antimatter asymmetry \cite{Bai:2013xga}, and the stabilization of the electroweak scale \cite{Chacko:2005pe, Burdman:2006tz,Craig:2015pha}.  More broadly, they represent a rather generic example of particle physics beyond the SM that could have flown under the radar of most existing searches at the LHC.  Signatures of such hidden valleys often feature non-isolated objects, high-multiplicity and/or soft final states, and long-lived species, and thus frequently represent a challenging target for existing analysis strategies at the LHC.  These ``dark shower'' signatures can give rise to spectacular and nonstandard final states, see e.g.~\cite{Strassler:2006im,Strassler:2008bv,Kang:2008ea,Harnik:2008ax,Cohen:2015toa,Schwaller:2015gea,Knapen:2016hky}, and are an emerging frontier for upcoming runs at the LHC \cite{Albouy:2022cin}.

A rigorous and comprehensive map of the theoretical possibilities within the dark shower framework is not yet possible.  However, some useful mileage can be gained by systematically classifying and constraining the portals through which dark hadrons may decay back to the SM \cite{Renner:2018fhh, Li:2019ulz, Knapen:2021eip,Bensalem:2021qtj}. Notably, this strategy has demonstrated that models featuring (sub-)GeV-scale dark mesons can only give rise to a high multiplicity of {\em visible} final particles in a subset of possible portal scenarios.
In particular, soft muonic final states are  well-motivated, especially if the dark states are short- or moderately long-lived ($c\tau \lesssim 1$ cm). If the event is sufficiently energetic, these soft muonic final states can be searched for inside (semi-visible) jets \cite{Pierce:2017taw,Cheng:2019yai,Zhang:2021orr,Cheng:2021kjg,Cazzaniga:2022hxl}. On the other hand, scouting techniques are the natural choice for scenarios where e.g.~the $H_T$, dijet or missing transverse energy (MET) triggers are not very effective.
 The sensitivity of LHCb in particular to low-mass resonances produced in dark showers has been noted earlier \cite{Pierce:2017taw,Cheng:2019yai,Aaij:2020ikh,Cheng:2021kjg}, as well potential additional sensitivity at Belle II \cite{Bernreuther:2022jlj}. In this work we take a close look at the complementarity of the recent CMS low-mass displaced dimuon search \cite{CMS:2021sch} with that at LHCb \cite{Aaij:2020ikh}, and show that both are powerful probes of hidden valley models in which the dark particles have 
tracker-scale lifetimes. We also suggest a number of extensions of these analyses, which could extend their discovery potential to hidden valley models with different vertex topologies.

We carry out this analysis in the context of a benchmark model that has two flavors of light dark quarks and an elementary dark photon. 
This benchmark model has features that are reminiscent of both neutral naturalness models on one hand \cite{Chacko:2005pe} and models of strongly-interacting massive particle (SIMP) dark matter with a  massive dark photon on the other \cite{Lee:2015gsa,Hochberg:2015vrg, Berlin:2018tvf,Katz:2020ywn}.
However, the primary aim of the model developed here is to serve as a useful reference for comparing and developing experimental sensitivities to dark shower events in the challenging low-mass regime. Our two-flavor dark photon model is designed to realize a variety of possible dimuon signatures with a relatively limited number of model parameters. It is thus proposed somewhat in the spirit of a simplified model approach to analysis design; however, we emphasize that it is a theoretically self-consistent, UV-complete model (as far as hadronization uncertainties allow).

The remainder of this paper is organized in two parts: the phenomenology and experimental sensitivity estimates are discussed in the bulk of the paper, while we reserve model-building aspects to the appendices.  In Sec.~\ref{sec:model} we summarize the essential features of our benchmark model; theorists interested in the model itself may find it beneficial to read the appendices here.  We describe our procedure for recasting the existing CMS and LHCb searches in Sec.~\ref{sec:analysis}
and present results in Sec.~\ref{sec:results}. Our conclusions and recommendations are presented in Sec.~\ref{sec:conclusions}. Appendix~\ref{app:appendix} contains further theoretical details of the benchmark model and its Pythia implementation; some additional figures can be found in Appendix~\ref{app:appendix_eff}.

\section{Benchmark dark shower model \label{sec:model}}

Here we develop a relatively simple benchmark model that can realize a range of distinct dimuon signatures depending on the choices made for its mass spectrum.  In this section we provide an overview of the model's collider phenomenology, with further model details given in Appendix~\ref{app:appendix}.

\subsection{Production}
Throughout this study, we assume that the dark shower is initiated through an exotic decay of the SM Higgs boson. This is theoretically the simplest production mode, as it does not require the introduction of a new, heavy mediator to initiate the dark shower. The SM Higgs ($h$) is moreover very narrow, making it a sensitive probe of beyond the Standard Model physics in general \cite{Curtin:2013fra,Cepeda:2021rql} and hidden valley models in particular \cite{Strassler:2006ri}. Studying dark shower production in exotic Higgs decays is especially well-motivated given the development of hidden valley models that aim to address the hierarchy problem, such as neutral naturalness models \cite{Chacko:2005pe, Burdman:2006tz,Craig:2015pha}.  From an experimental point of view, Higgs production is also one of the more challenging possibilities owing to the relatively low-$p_T$ final states that it generates.  For our present purposes, this makes exotic Higgs decays an excellent benchmark scenario to illustrate the power of data scouting techniques, as alternative trigger paths for these events would need to rely on associated SM objects produced along with the Higgs in vector boson fusion or associated $VH$ production, with accordingly smaller production cross sections.

The Higgs boson initiates dark showers by decaying into pairs of dark quarks through an interaction of the form $h \bar q_D q_D$.  This interaction can be readily constructed in the benchmark model described in detail in App.~\ref{app:quarksector}; in this benchmark, moreover, the size of the dark quark couplings to the SM Higgs can  be adjusted without significant impact on the dark hadron masses and lifetimes.
For our purposes, the branching ratio $\text{Br}(h\to \text{HV})$ into dark showers is the primary quantity of interest, and can be treated as a free parameter.  We will therefore present our results in terms of this branching ratio.

\begin{figure}\centering
\includegraphics[width=0.35\textwidth]{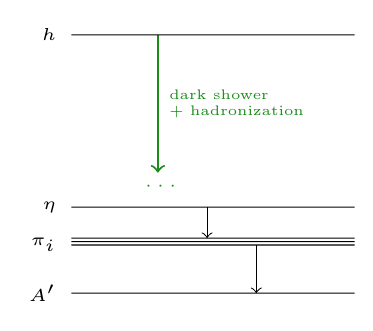}\hspace{2cm}
\caption{Schematic representation of particle content and decay chains in the two-flavor hidden valley benchmark model developed in this work. The lowest-lying dark hadrons are an approximate dark-isospin triplet $(\pi_1,\pi_2,\pi_3)$ and heavier pseudoscalar $\eta$. 
The $\cdots$ represent  heavier mesons and baryons. The green arrow represents the dark parton shower and hadronization steps, which result in multiple $\eta$s and $\pi_i$s per event. Finally, the dark mesons couple to an elementary dark photon $A'$, which controls the (displaced) decays of the $\eta$ and/or some of the $\pi_i$ to final states with SM fermions, as detailed in Sec.~\ref{sec:vertextopology}
\label{fig:model}}
\end{figure}

\subsection{Dark sector spectrum\label{sec:showerhadronization}}

Once the Higgs has decayed to a pair of dark quarks, those quarks will undergo showering and hadronization within the dark sector, which we model with the hidden valley module \cite{Carloni:2010tw,Carloni:2011kk} in Pythia 8 \cite{Sjostrand:2014zea}. The hidden valley module recently received a major update expanding its capabilities to describe dark meson multiplets with broken flavor symmetries \cite{Albouy:2022cin}, which was essential to  implement the model we develop in this work. 

We fix the number of colors and flavors to $N_c=3$ and $N_f=2$ respectively, and allow for the running of the dark gauge coupling. The two-flavor case is the most minimal setup that generates the range of displaced vertex topologies that we study in Sec.~\ref{sec:vertextopology}. We further add an additional, massive $U(1)$ interaction to facilitate the various decay chains described below. This $U(1)$ is spontaneously broken, and the resulting massive dark photon kinetically mixes with the SM photon to supply a muon-philic decay portal back to the SM. Fig.~\ref{fig:model} shows a schematic representation of the spectrum and the most important decay chains; the model is defined in full detail in Appendix~\ref{app:appendix}.

\begin{figure}
\includegraphics[width=0.48\textwidth]{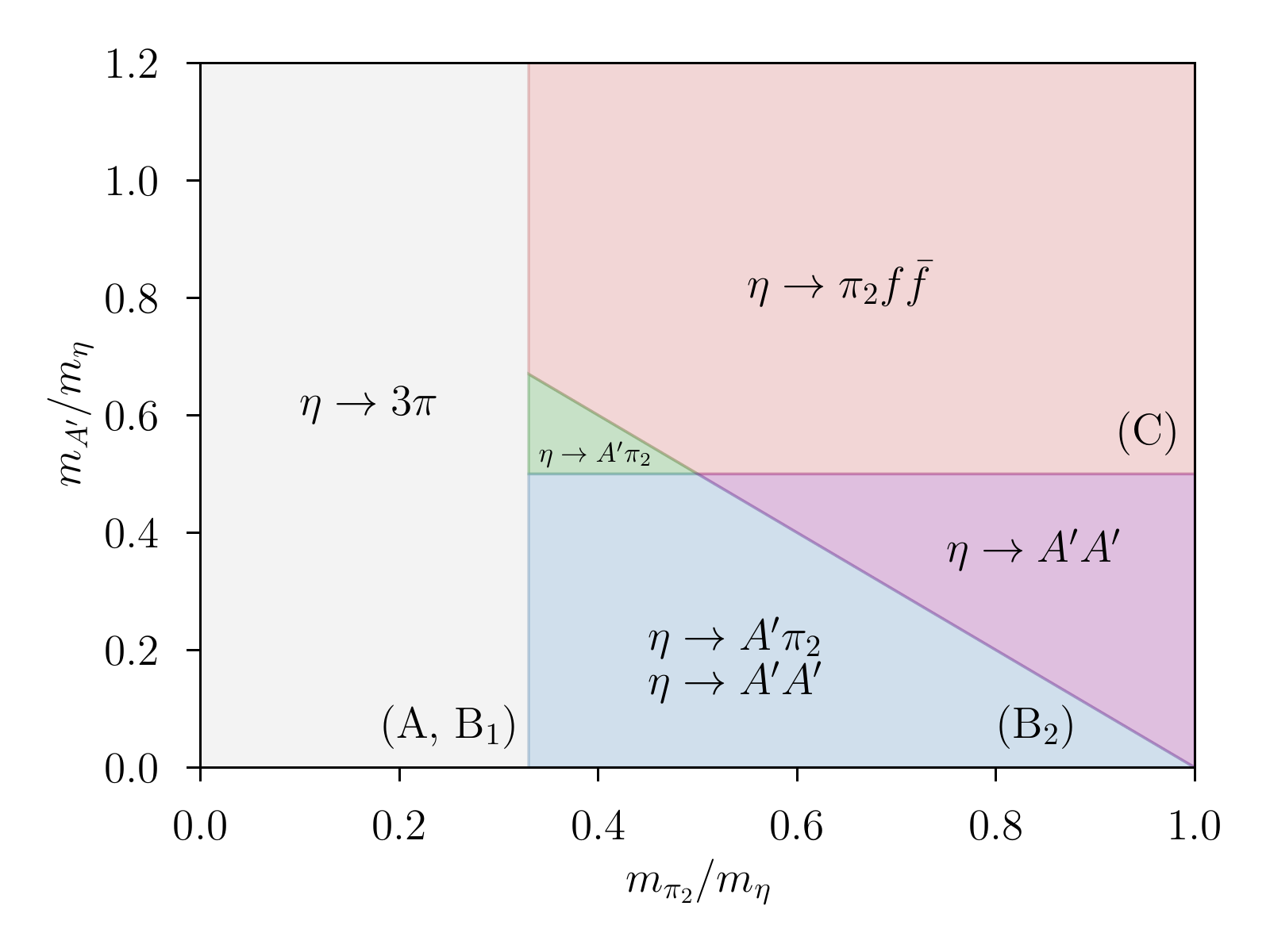}
\caption{Summary of the $\eta$ phase space and available decay chains. The $\eta\to 3\pi$ decay (gray region) is always prompt, while the three-body decay  $\eta \to \pi_2 \bar f f$ (red) tends to be displaced, with $f\bar f$ representing SM final states. The remaining modes can be either prompt or displaced, depending on the choice of model parameters. If the $\eta$ is taken to be long-lived in the blue region, the $\eta \to A' \pi_2$ mode always dominates. The (A), (B1), (B2) and (C) labels refer to the decay topologies in Sec.~\ref{sec:vertextopology}.\label{fig:decays} }
\end{figure}

In a bit more detail, we assume that the confining dark gauge group dynamically breaks the hidden sector $SU(2)\times SU(2)$ flavor symmetry down to its diagonal subgroup, as is the case in the SM pion sector. This leads to three pseudo-goldstone bosons, which we label $\pi_1, \pi_2$ and $\pi_3$. There is also a heavier $\eta$ meson, which corresponds to the would-be goldstone boson of the anomalous axial $U(1)$. We further assume that the couplings of $A'$ to dark sector quarks break parity, charge conjugation, and isospin. This isospin breaking will lead to small mass splittings between the various $\pi_i$, which are not relevant for the phenomenology we are interested in here. The charges of the $A'$ are however chosen such that the following interaction is non-vanishing:
\begin{equation}
\mathcal{L} \supset g \sin \theta A'^\mu \left(\pi_2 \partial_\mu \eta - \eta \partial_\mu \pi_2  \right),
\end{equation}
where $g$ is the $U(1)$ gauge coupling and $\sin\theta$ a mixing angle parameterizing the isospin violation. (See Appendix~\ref{app:mesonsector} for details). This interaction enables the decay $\eta \to A'\pi_2$. Both the $\pi_3$ and $\eta$ can also decay to a pair of $A'$ through the dark sector chiral anomaly, provided this channel is kinematically open. Meanwhile, the $\pi_1$ and $\pi_2$ are stable on collider timescales and contribute MET to the event. 

Since the $\eta$ is heavier than the $\pi_i$, one expects hidden sector hadronization to produce it less frequently. Hadronization, as always, is one of the major sources of theoretical uncertainty in hidden valley models.  In particular,  the relative probability of producing different particle species in dark hadronization cannot be calculated from first principles, and thus the choices made in the hadronization component of the Pythia HV module must also be considered as part of the definition of any benchmark model; for an up-to-date discussion of these choices see \cite{Albouy:2022cin}.  
The Pythia cards for our model points are generated with our public python script \cite{code}; more detail about the Pythia implementation can be found in Appendix \ref{sec:pythiacard}.

The available decay channels for the $\eta$ are shown schematically in Fig.~\ref{fig:decays}. If \mbox{$m_{\eta}> 3 m_{\pi}$}, the prompt decay \mbox{$\eta\to 3 \pi$} will always dominate. This case leads to a marginally enhanced  multiplicity of $\pi$'s, where the $\pi_3$ decays as $
\pi_3 \to A'A'\to \mathrm{SM}$. 
The \mbox{$\eta\to A'A'$} decay mode is only open for \mbox{$m_\eta>2 m_{A'}$}, while \mbox{$\eta \to A' \pi_2$} may occur if \mbox{$m_\eta> m_{A'}+m_{\pi_2}$}. If both channels are available (blue region in  Fig.~\ref{fig:decays}), the $\eta \to A' \pi_2$ channel dominates, especially if the $\eta$ is long-lived. Finally, if none of these conditions are satisfied, the $\eta$ can decay through an off-shell $A'$ into a SM final state plus $\pi_2$. One expects this decay to occur with a proper decay length of $\mathcal{O}(\mathrm{cm})$ or larger. We refer to Appendix~\ref{app:branchingratios} for more details.

\subsection{Vertex topologies\label{sec:vertextopology}}
\begin{figure}
\includegraphics[height=4.cm]{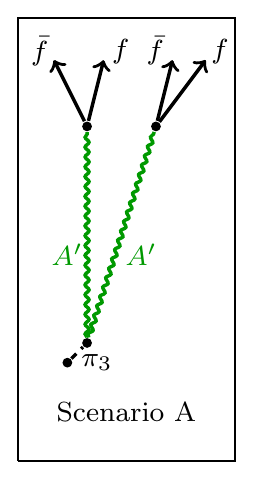}\hspace{-0.2cm}
\includegraphics[height=4.cm]{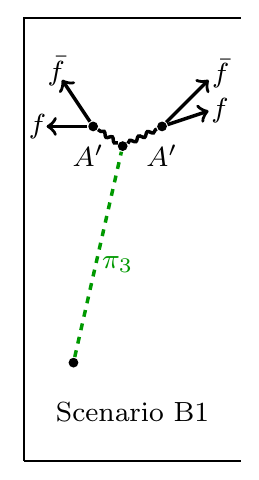}\hspace{-0.45cm}
\includegraphics[height=4.cm]{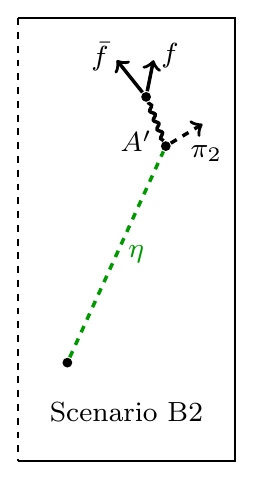}\hspace{-0.2cm}
\includegraphics[height=4.cm]{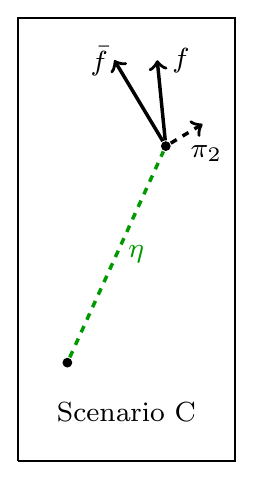}
\caption{Schematic representation of the decay topologies considered in this work, with the $f$ ($\bar f$) SM (anti-)fermions, e.g. $\mu^-$ ($\mu^+$). The green lines indicate macroscopic decay lengths. \label{fig:topology diagrams}}
\end{figure}

Our benchmark model can generate three possible vertex topologies, as summarized in Fig.~\ref{fig:topology diagrams}. Concretely, we consider the following scenarios:
\begin{itemize}
\item \textbf{Scenario A (resonant, pointing):} This scenario is realized when the $\eta \to 3\pi$ channel is open (gray region in Fig.~\ref{fig:decays}) such that the $\pi_3$ meson is the only dark sector meson that decays directly to the SM.  
We assume that this occurs promptly, through the $\pi_3\to A' A'$ channel, while the $A'$ decays to the SM by kinetically mixing with the SM photon. If this mixing angle is small, $A'$ can however be long-lived, and we treat its lifetime ($c\tau_{A'}$) as a free parameter. This scenario produces resonant dimuon pairs, for which the vector sum of their momenta points back to the beamline.

\item \textbf{Scenario B (resonant, non-pointing):} This case occurs when on-shell dark photons are produced in the displaced decay of a long-lived parent particle.  The vector sum of the momentum of the resulting resonant dimuon pairs does not point back to the beamline. We realize this in two different ways in our benchmark model: 

{\bf Scenario B1 (higher visible multiplicity):} In this case we keep the $\eta \to 3\pi$ and $\pi_3\to A' A'$ decay channels open, but choose the dark gauge coupling small enough that the $\pi_3\to A' A'$ decay becomes displaced (see Appendix~\ref{app:mesonsector}).   For a given meson mass spectrum, this case has the same average multiplicity of displaced dimuon pairs as Scenario A.  When the dark photon lifetime is taken to be prompt, the dimuon pairs are produced at the same vertex as the visible particles that come from the decay of the other dark photon. 

{\bf Scenario B2 (lower visible multiplicity):} In this scenario, we instead take the $\eta \to 3\pi$ and $\pi_3\to A'A'$ channels to be kinematically closed, while $\eta\to \pi_2 A'$ is allowed (blue and green regions in Fig.~\ref{fig:decays}). We further assume that the $\eta$ is long-lived, but that the $A'$ decays promptly to the SM, again by mixing with the SM photon. We treat the $\eta$ lifetime ($c\tau_\eta$) as a free parameter.  In this scenario, all three dark pions ($\pi_{1,2,3}$) now escape the detector, so the dimuon pairs are the only visible particles associated with the displaced vertex.  In this case the average dimuon multiplicity depends on the average $\eta$ multiplicity $\langle N_\eta\rangle$, and is substantially smaller than in Scenario A.

\item \textbf{Scenario C (non-resonant):} This scenario holds when the $\eta$ can only decay through the three-body process involving an off-shell dark photon $\eta \to \pi_2 f\bar f$ (red region in Fig.~\ref{fig:decays}). The $\eta$ is typically long-lived in this case (see Appendix~\ref{app:branchingratios}). This scenario produces non-resonant dimuon pairs, with the same dependence on $\langle N_\eta\rangle$ as in Scenario B2.\footnote{Displaced vertices with non-resonant muon pairs can also be realized for $m_\eta > 3 m_\pi$ and $2m_{A'}>m_{\pi_3}$, in which case $\pi_3$ has three-body decays $\pi_3\to A' f\bar f$; however in this case the $\pi_3$ proper lifetime tends to be prohibitively long in the kinematic regime relevant for scouting searches.} In this scenario all dark pions ($\pi_{1,2,3}$) are again  detector-stable.
\end{itemize}

As we will see, the existing CMS and LHCb searches already have excellent sensitivity for scenario A, but the CMS search is not currently targeting scenario B. Neither search currently targets scenario C. 

\subsection{Parameter choices and hadron multiplicities}

Our $N_f=2$ confining dark sector has a substantial number of parameters, some of which are related to one another by chiral perturbation theory, as worked out in Appendices \ref{app:quarksector} and \ref{app:mesonsector}. We choose the following parameters as our maximal independent set: the number of dark sector colors ($N_c$), the mass of $\pi_2$ ($m_{\pi_2}$), the mass of the $\eta$ ($m_\eta$), the dark sector confinement scale ($\Lambda$), the scale of chiral symmetry breaking ($f$), the degree of isospin breaking $(\sin\theta)$, the dark photon mass ($m_{A'}$), the dark $U(1)$ gauge coupling ($g$) and the mixing parameter of the $A'$ with the SM photon ($\epsilon$). The mass spectrum and decay tables of the dark sector $\pi_i$, $\eta$ and $A'$ are fully specified in terms of these parameters, as provided in the appendices and by our python tool \cite{code}.

 We follow Pythia's definition of the confinement scale $\Lambda$, where it is given by the dimensional transmutation parameter of the one-loop $\beta$-function of the dark sector $SU(N_c)$ coupling.\footnote{While this work was in the final stages of completion, Pythia version 8.309 was released, which now also includes the two- and three-loop $SU(N_c)$ $\beta$-functions. All our results were obtained with Pythia version 8.308.} One may expect this scale to roughly correspond to the scale at which chiral perturbation theory ceases to be reliable, in other words
\begin{equation}\label{eq:approxQCD}
4\pi f\approx \Lambda.
\end{equation}
For definitiveness, we assume this relation to hold exactly; changing this assumption would lead to $\mathcal{O}(1)$ variations in the decay widths of some of the dark mesons. In addition, our experience with the SM suggests that the meson associated with the anomalous axial $U(1)$ flavor symmetry has a mass comparable to the confinement scale. In our model, this role is played by the $\eta$-meson and as such we expect
\begin{equation}\label{eq:approxeta}
m_{\eta}\approx \Lambda.
\end{equation}
This expectation famously does not hold for theories in the large-$N_c$ limit  \cite{Witten:1980sp}. We will take \eqref{eq:approxeta} to hold as an equality as well and choose $N_c=3$. We further take $\sin\theta =0.1$, such that isospin breaking is relatively small and
\begin{equation}
m_{\pi_1}\approx m_{\pi_2} \approx m_{\pi_3}
\end{equation}
for all our benchmark points.  Choosing a small value of the isospin-breaking parameter $\sin \theta$ is also important for ensuring that the Pythia hidden valley module provides a reasonable approximation to the hadronization process, as we detail in Appendix~\ref{sec:pythiacard}. 

The parameters $g$ and $\epsilon$ on the other hand are responsible for setting the lifetimes of the $A'$, $\pi_3$ and $\eta$. As such, all our benchmark points in this paper only differ by different choices for the masses $m_{\pi_3}$, $m_{\eta}$ and $m_{A'}$, and the choices for the lifetimes of the corresponding particles ($c\tau_{\pi_3}$, $c\tau_{\eta}$ and $c\tau_{A'}$). Concretely, the individual scenarios are defined by choosing
\begin{itemize}
\item \textbf{Scenario A:} $c\tau_{\pi_3}=c\tau_{\eta}=0$, \mbox{$m_\eta> 3 m_{\pi_2}$}
\item \textbf{Scenario B1:} $c\tau_{A'}=c\tau_{\eta}=0$, \mbox{$m_\eta> 3 m_{\pi_2}$}
\item \textbf{Scenario B2:} $c\tau_{A'}=0$,  
\mbox{$m_\eta> m_{\pi_2}+m_{A'}$},\\ $\pi_3$ detector-stable
\item \textbf{Scenario C:} $c\tau_{A'}=0$,
\mbox{$m_\eta< m_{\pi_2}+m_{A'}$},\\$\pi_3$ detector-stable
\end{itemize}
where 
``zero'' $c\tau$ is meant as shorthand for ``promptly decaying'', which is consistent with experimental constraints on visible dark photons for $m_{A'}\gtrsim10$ MeV \cite{Gori:2022vri}.
The remaining parameters are specified in the relevant figures.

\begin{figure}
\includegraphics[width=0.48\textwidth]{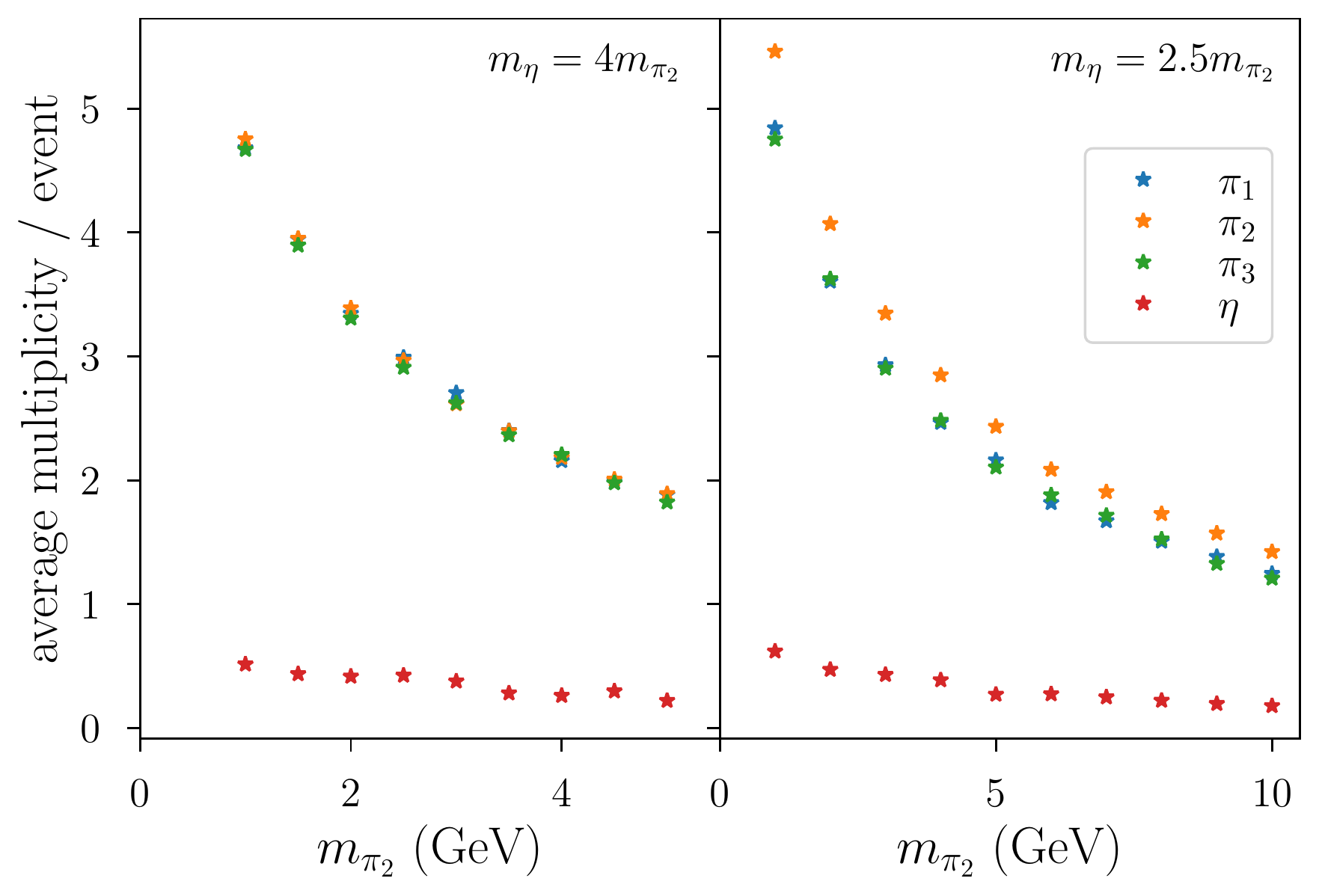}
\caption{Meson multiplicities for two representative choices of the $m_{\eta}/m_{\pi_2}$ mass ratio. The left panel corresponds to scenarios A and B1, while the right panel corresponds to scenarios B2 and C.\label{fig:multiplicities}}
\end{figure}

With these assumptions, we can investigate average particle multiplicities, as generated by the Pythia 8 hidden valley module. We show the average meson multiplicities in Fig.~\ref{fig:multiplicities} as a function of pion mass for two example choices of the mass ratio $m_{\eta}/m_{\pi_2}$. Naturally, the multiplicity of all meson species rises as the meson mass and thus the confinement scale are lowered. We furthermore see that the $\pi_i$ multiplicity is a factor of several higher than the $\eta$ multiplicity. This is in part due to the lower mass of the $\pi_i$, which means it is more likely to be produced in the hadronization model. Moreover, the dark sector isospin triplet vector mesons, the analogues of the $\rho$-mesons in the SM, are assumed to decay promptly to the $\pi_i$, further enhancing their multiplicity.

At the level of the meson mass spectrum, the difference between scenarios (A, B1) vs. scenarios (B2, C) is whether the $\eta\to 3 \pi$ channel is kinematically open. The former case comes with a slightly higher pion multiplicity. In scenarios B2 and C, the $\eta\to A' \pi_2$ decay mode explicitly breaks isospin, and is the reason why the $\pi_2$ multiplicity is  higher than the $\pi_{1,3}$ multiplicities in the right-hand panel of Fig.~\ref{fig:multiplicities}.

\section{Analysis}
\label{sec:analysis}

We consider two searches for low-mass displaced dimuon resonances from LHCb \cite{Aaij:2020ikh} and  CMS  \cite{CMS:2021sch}.  Both analyses are powerful probes of the dark shower signatures presented in the previous section. Here we briefly describe both searches, as well as our procedures for event generation and recasting.

\subsection{CMS displaced dimuon search\label{sec:existing}}

The recent CMS search \cite{CMS:2021sch} records displaced low mass dimuon pairs with $p_T$ thresholds as low as $\sim 3$ GeV, at a rate of 3 kHz. This large rate is feasible because CMS only records a very small fraction of the total event information, which includes the muon four-momenta, number of hits per muon track, muon isolation, and track-quality information. The search used 101 $\text{fb}^{-1}$ of data at 13 TeV.

The most important selection criteria in the CMS search are listed in Tab.~\ref{tab:selection}. These include cuts on the transverse momentum ($p_T$) and pseudorapidity ($\eta$) of the muons, as well as a cut on the distance between the dimuon vertex and the beamline ($L_{xy}$). In addition CMS also requires the vector sum of the muon momentum vectors to point back to the beamline in order to suppress backgrounds from fake vertices. Concretely, they impose a hard cut on the azimuthal angle between the vector sum of the muon momenta and the vector connecting the beamline to the displaced vertex. We will call this variable $\Delta \phi(\sum \vec p_T(\mu),\vec x)$, with $\vec x$ representing the location of the displaced vertex in the detector frame. This pointing requirement is automatically satisfied at truth-level for scenario A in Sec.~\ref{sec:vertextopology}, but not for scenarios B and C. We will speculate on loosening this restriction in Sec.~\ref{sec:scenarioB}. CMS also imposes  further selection criteria, such as track and vertex quality requirements; we refer to \cite{CMS:2021sch} for more details. In addition to this set of baseline cuts, CMS defines a set of increasingly restrictive signal regions, the most relevant of which are listed in Tab.~\ref{tab:selection}. In particular, an event is put in the ``isolated'' category if the scalar sum of the $p_T$ of all tracks within a cone of $\Delta R<0.3$ around each muon does not exceed 20\% of the $p_T$ of the muon. 

CMS provides trigger and offline efficiencies for each signal region, as well as the number of excluded events. We use this data in our re-interpretation, as described in Sec.~\ref{sec:simulation}. The collaboration evaluates signal efficiencies for two well-motivated benchmark topologies: 
an exotic $B$-meson and exotic Higgs decay, respectively $B\to X_s S$ and $h\to A'A'$, where the scalar ($S$) and vector $(A')$ are assumed to decay to $\mu^+\mu^-$. We validated our reinterpretation procedure by simulating both signal models and verifying that for each model the exclusions we estimate by applying the provided efficiencies are in excellent agreement with the experimental results. Since we expect the muon $p_T$ spectrum to be rather soft for most low-mass hidden valley models, we consider $B$-meson decay to be the more appropriate comparison model.

\begin{table*}[t]
\begin{tabular}{c|c|c}
& \textbf{fiducial cuts} & \textbf{signal regions}\\\hline
\multirow{6}{*}{\textbf{CMS}}  & $p_{T}(\mu)> 3$ GeV&0.0 cm $<L_{xy}$ (w/ and w/o isolation)\\
  & $|\eta(\mu)|<2.4$&0.2 cm $<L_{xy}$ (w/ and w/o isolation)\\
  &$L_{xy}<$ 11 cm &1.0 cm $<L_{xy}$ (w/ and w/o isolation)\\
  &$\Delta \phi(\sum \vec p_T(\mu),\vec x)<0.02$ &2.4 cm $<L_{xy}$ (w/ and w/o isolation)\\
  & &3.0 cm $<L_{xy}$  (w/ and w/o isolation)\\
  & &7.0 cm $<L_{xy}$  (w/ and w/o isolation)\\\hline
   \multirow{5}{*}{\textbf{LHCb}} &$p_{T}(\mu)> 0.5$ GeV, $|\vec p(\mu)|> 10$ GeV&2 GeV $<p_T(A')<3$ GeV (w/ and w/o pointing)\\
 &$2<\eta(\mu)<4.5$& 3 GeV $<p_T(A')<5$ GeV (w/ and w/o pointing)\\
 &$\sqrt{p_T(\mu^+)p_T(\mu^-)}>$ 1.5 GeV&5 GeV $<p_T(A')<10$ GeV (w/ and w/o pointing)\\
  &1.2 cm $< L_{xy}<$ 3 cm \\
  &$\alpha(\mu^+,\mu^-)>$ 3 mrad \\
\end{tabular}
\caption{Most important fiducial cuts for the CMS \cite{CMS:2021sch} and LHCb \cite{Aaij:2020ikh} searches, as well as most relevant signal regions for the purpose of the study in this work. The CMS signal regions are non-exclusive, while the LHCb signal regions are exclusive. \label{tab:selection}}
\end{table*}

\subsection{LHCb displaced dimuon search}
The LHCb detector is specifically optimized to search for low mass resonances and has consistently been on the forefront of the usage of online analysis strategies. This has resulted in excellent sensitivity to light beyond-the-SM particles, in particular if these particles have a substantial branching ratio to muons \cite{LHCb:2015nkv,Aaij:2016qsm,Aaij:2019bvg,Aaij:2020ikh}. Here we focus on LHCb's recent inclusive search for displaced dimuon resonances \cite{Aaij:2020ikh}, which was carried out using 5.1 $\textrm{fb}^{-1}$ of data at 13 TeV. The most important fiducial cuts are listed in Tab.~\ref{tab:selection}. The main differences with respect to the CMS selection are: \emph{i)} a substantially lower $p_T$ cut on the muons; \emph{ii)} the more forward acceptance of the LHCb detector; \emph{iii)} the smaller $L_{xy}$ range; and \emph{iv)} a loose cut on the opening angle between the two muons $\alpha(\mu^+,\mu^-)$.

\subsection{Simulation framework\label{sec:simulation}}

For our signal Monte Carlo samples, we generated Higgs bosons through the gluon fusion channel using Pythia 8.308 \cite{Sjostrand:2014zea} and implemented the decay mode to dark quarks. The dark quarks are subsequently showered and hadronized with the Pythia 8 hidden valley module \cite{Carloni:2010tw,Carloni:2011kk}. Further detail on the Pythia implementation of our benchmark model is given in Appendix~\ref{sec:pythiacard}. 
All signal samples were generated with promptly-decaying particles and the lifetime dependence of the acceptance was accounted for by the reweighting procedure described in \cite{Knapen:2022afb}. Concretely, for each event, we identify all vertices that pass the selection cuts in Tab.~\ref{tab:selection}, except for the $L_{xy}$ requirements. For the CMS analysis, we moreover compute the isolation criteria on a vertex-by-vertex basis and classify events accordingly. We work with the truth-level four-momenta of the muons, effectively neglecting smearing due to the finite detector resolution. Reconstruction and trigger efficiencies were however incorporated to the extent possible, as described below. 
 
 To efficiently account for the cuts on $L_{xy}$, we assign an acceptance weight $\wacc (v)$ to each vertex $v$ that passed these cuts:
\begin{equation}
\wacc (v)=e^{-\frac{L_{xy}^-}{\beta\gamma\cosh(\eta)c\tau}}-e^{-\frac{L_{xy}^+}{\beta\gamma\cosh(\eta)c\tau}},
\end{equation}
which corresponds to the probability that the long-lived particle corresponding to $v$ decays within two co-axial cylinders with inner and outer radii $L_{xy}^-$ and $L_{xy}^+$. In other words, here the values $[L_{xy}^-,L_{xy}^+]$ represent the edges of the $L_{xy}$ bins used in the analysis.  Here $c\tau$, $\beta\gamma$, and $\eta$ are respectively the proper lifetime,  boost, and pseudorapidity of the long-lived particle. The full weight of the vertex is obtained after multiplying $\wacc(v)$ by the probabilities that the vertex would pass the trigger and offline selections, denoted by $\etrig(v)$ and $\eoff(v)$, or
\begin{equation}
w(v)\equiv \wacc (v) \times \etrig(v)\times \eoff(v).
\end{equation}
For the CMS analysis, $\etrig(v)$ and $\eoff(v)$ are provided in the supplementary material attached to the analysis, for the signal regions listed in Tab.~\ref{tab:selection}. For the LHCb analysis we assume $\etrig(v)\approx \eoff(v)\approx 1$, in line with the study in \cite{Ilten:2015hya}.

For searches requiring just a single vertex, we can simply define the weight of the whole event $w(e)$ as 
\begin{align}
w(e)&\equiv 1- \prod_{v\in \{\mathrm{vertices}\}} (1-w(v)) \\
&\approx \sum_{v\in \{\mathrm{vertices}\}} w(v)
\end{align}
where the approximation is justified whenever $w(v)\ll 1$ for all vertices. For the CMS analysis, we will also consider a signal region where at least \emph{two} vertices are reconstructed in the event. In this case, the event weight is defined as
\begin{align}
w(e)\equiv& 1-\!\!\! \prod_{v\in \{\mathrm{vertices}\}}\!\!\! (1-w(v)) \nonumber \\
&- \!\!\!\sum_{v\in \{\mathrm{vertices}\}}\!\!\! w(v)\!\!\! \prod_{v'\in \{\mathrm{vertices}\}\setminus \{v\}} \!\!\! (1-w(v')).
\end{align}
This reweighting strategy allows us to efficiently compute limits for arbitrary values of $c\tau$, without the need to regenerate the Monte Carlo samples.

The procedure outlined above gives us the signal efficiency as a function of the model parameters. We show the signal efficiency for select benchmark points in Fig.~\ref{fig:eff_appendix} of Appendix~\ref{app:appendix_eff}.
We can directly compare this with the number of excluded events in the case of CMS \cite{CMS:2021sch} or with the limit on the fiducial cross section for LHCb \cite{Aaij:2020ikh} and extract a bound on the branching ratio of the SM Higgs to Hidden Valley model under consideration for each signal region. CMS and LHCb present their limits binned in terms of $L_{xy}$ and $p_T$ respectively (see Tab.~\ref{tab:selection}). In both cases we take the bin with the strongest limit to represent our limit. CMS only reports results for dimuon pairs whose reconstructed momentum satisfies the pointing cut in Tab.~\ref{tab:selection}.
For LHCb, we use the pointing selection for Scenario A and the non-pointing selection for Scenarios B1 and B2. 

\section{Results}
\label{sec:results}

\begin{figure}
\includegraphics[width=0.48\textwidth]{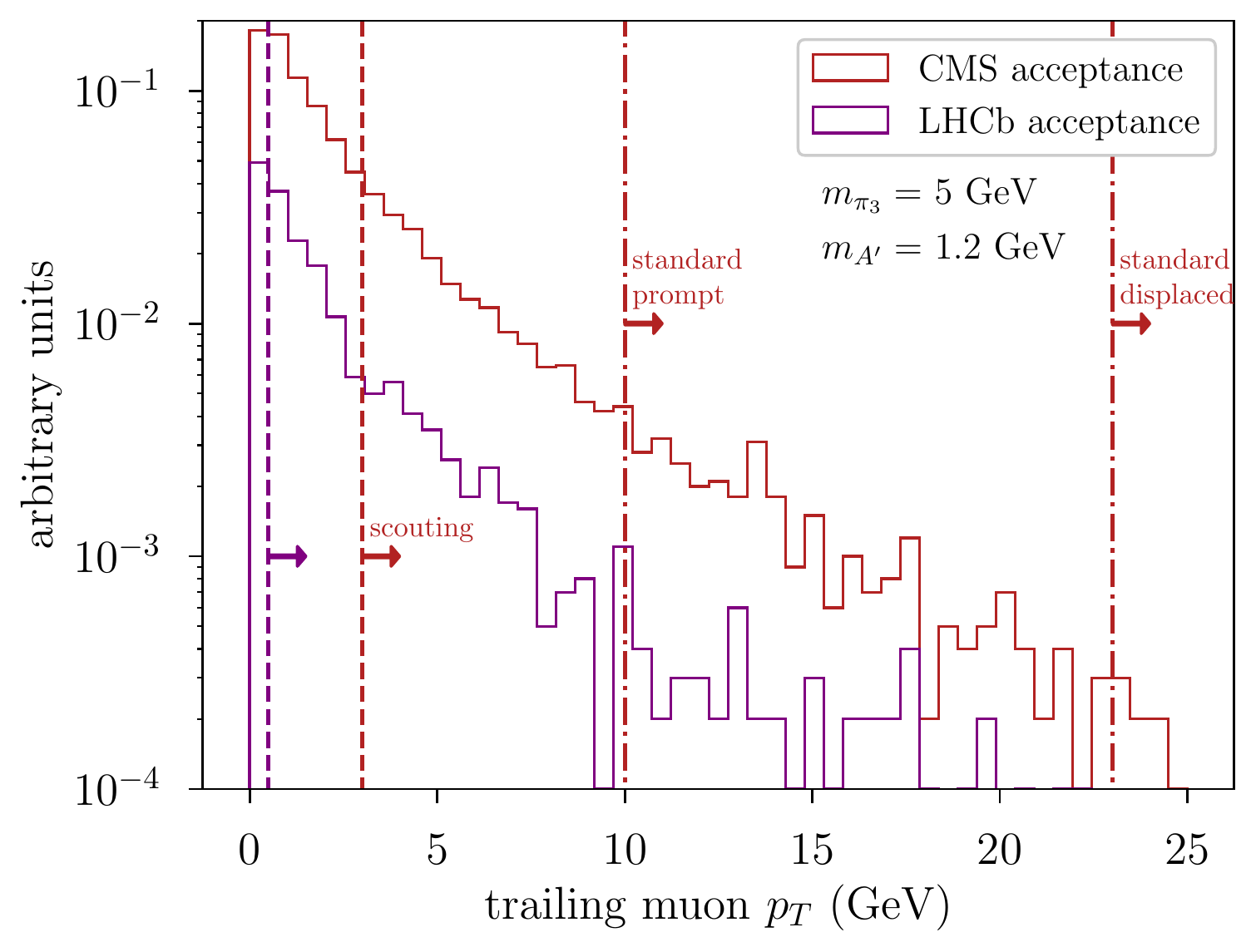}
\caption{The $p_T$ spectra of the trailing muon in each vertex for an example Scenario A model point, within pseudorapidity ranges covered by the CMS tracker and LHCb VELO detectors (red and purple respectively). Dashed lines indicate the $p_T$ threshold of the corresponding analyses, see Sec.~\ref{sec:existing} for details. The dot-dashed lines indicate the threshold for the prompt and displaced dimuon triggers used in respectively \cite{CMS:2022qej} and \cite{CMS:2021pcy, ATLAS:2023vxg}.  The Pythia 8 card used to generate the events in this figure is included in Appendix~\ref{sec:pythiacard}.
\label{fig:pTplot}}
\end{figure}

The relative advantage of a scouting trigger is best seen by looking at the $p_T$ spectrum of the softer of the two muons in each vertex, as shown in Fig.~\ref{fig:pTplot} for an example working point. While signal events may be muon-rich, their muons are generally soft, and, depending on the parent particle lifetime, very often displaced.  The $p_T$ treshholds for displaced muon triggers are prohibitively high for this signal, as Fig.~\ref{fig:pTplot} demonstrates, while the scouting trigger can record events with trailing muon $p_T$ as low as \mbox{3.5 GeV}.  Thus the scouting trigger is uniquely suited to capture the regime where signals yield soft, displaced muon pairs. 

For muon pairs with a sufficiently small displacement to be picked up by prompt dilepton triggers, a softer trailing muon $p_T$ cut of 10 GeV is more reflective of ATLAS and CMS capabilities (e.g., \cite{CMS:2021pcy, ATLAS:2023vxg}).  But even this $p_T$ threshold still yields an acceptance of only a few percent for the model point in Fig.~\ref{fig:pTplot}. For the low c$\tau$/prompt regime, the scouting analysis in \cite{CMS-PAS-EXO-21-005} is therefore very pertinent.  While  the scouting trigger has much better signal acceptance, it also collects more background events than the traditional dimuon trigger, in particular for muon pairs produced at low to no displacement. In some models, the signal may contain additional handles such as jet substructure information or  hadronic displaced vertices that could help with background discrimination, but which are not retained in the scouting stream. In such cases, a fully offline analysis may prove to be more powerful. For the benchmark scenarios that we consider in this paper, we do not expect such handles to add much discriminating power in the regime where all dark decays are prompt. We thus suspect that even in the limit of prompt decays the scouting analysis in \cite{CMS-PAS-EXO-21-005} would outperform the standard dimuon trigger, although this statement is model-dependent. We leave a quantitative study of scouting for dark showers in the prompt regime to future work.

\subsection{Scenario A}

\begin{figure*}[t]
\includegraphics[width=\textwidth]{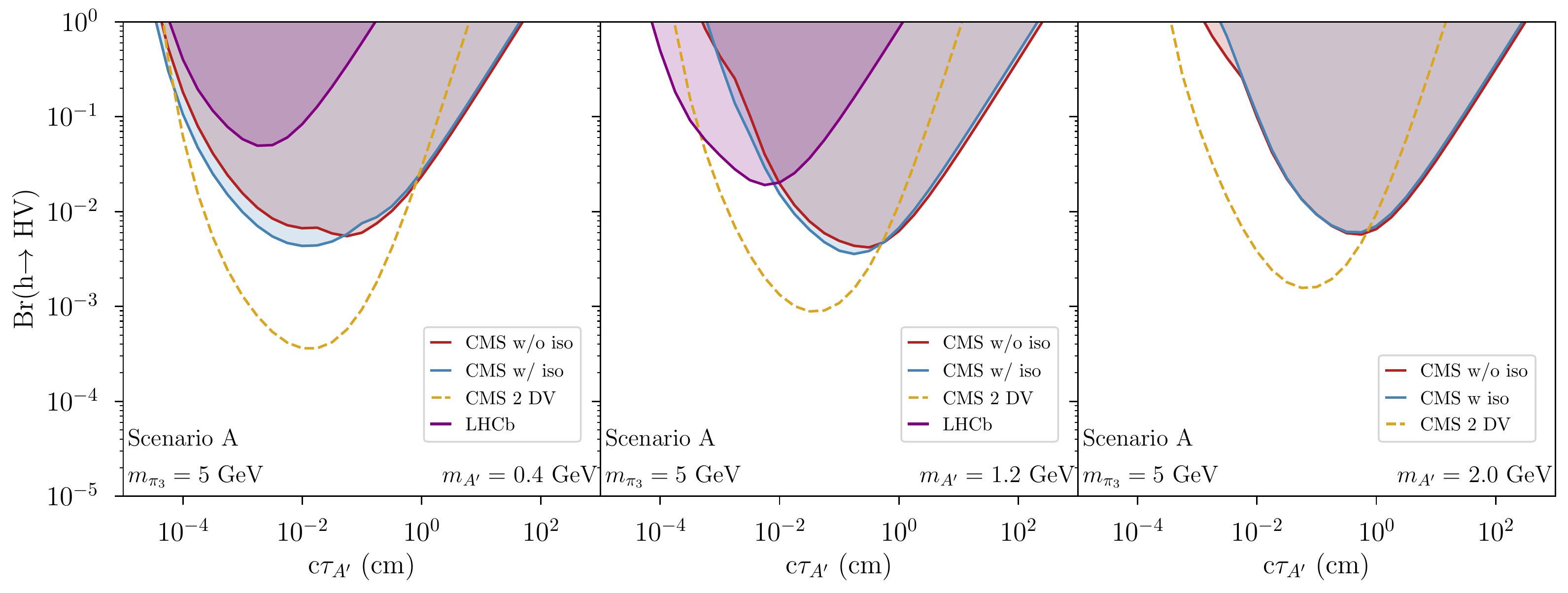}
\caption{ Limits on the branching ratio for the SM Higgs decaying to the hidden valley model described in Sec.~\ref{sec:model}, with $\Lambda=m_{\eta}=4 m_{\pi_3}$, $\sin\theta=0.1$ and $g=0.05$. (See Appendix~\ref{app:mesonsector} for details.) The bounds are shown as a function of the lifetime of the dark photon $c\tau_{A'}$, from our recast of the CMS dimuon scouting analysis \cite{CMS:2021sch} and the LHCb low mass dimuon analysis \cite{Aaij:2020ikh}. The dashed yellow line indicates the limit set by \cite{CMS:2021sch} when requiring two dimuon vertices per event, under the assumption that this suffices to suppress the background to negligible levels. (See text for details.)   \label{fig:ctauplot}}
\end{figure*}

We first discuss Scenario A, which produces resonant dimuon pairs whose reconstructed momenta point back to the beamline.  The bounds we obtain from both LHCb and CMS for benchmark dark shower models produced in exotic Higgs decays is shown in Fig.~\ref{fig:ctauplot} for a few example mass points. The blue and red shaded regions correspond to the bounds set by the CMS analysis with and without imposing isolation, respectively. The isolation efficiency for these signal benchmarks is roughly 40\%, which reflects that in this scenario the dark photons are produced in pairs.  We see that imposing isolation is somewhat beneficial for relatively low $m_{A'}$, where the SM backgrounds are substantial. For \mbox{$1\; \mathrm{GeV} \gtrsim m_{A'}\gtrsim 3 $ GeV} both selections perform comparably. For \mbox{$m_{A'}\gtrsim 3 $ GeV}, however, the background drops off sharply \cite{CMS:2021sch}, and the more inclusive selection sets the best limit.  The best sensitivity of the LHCb analysis is at somewhat lower $c\tau_{A'}$, as the dark photons in the LHCb acceptance tend to be more boosted than those in the CMS acceptance. Despite its much lower luminosity, the LHCb analysis outperforms the CMS displaced analysis for \mbox{$m_{A'}\gtrsim 1$ GeV} and $c\tau_{A'}\lesssim 1$ mm.\footnote{The CMS prompt scouting search \cite{CMS-PAS-EXO-21-005} appeared when our manuscript was in its final stage of completion. It likely has some sensitivity in the low $c\tau$ regime; we defer a detailed analysis of its reach for dark shower models to future work.}

One of the key features of dark shower topologies is that they tend to produce multiple long-lived particles per event. If the proper decay length of these long-lived particles is comparable to or smaller than the size of the detector and if the vertex reconstruction efficiency is sufficiently high, one expects that requiring an additional displaced dimuon vertex would be a powerful handle to reduce the backgrounds. To illustrate this, we derive a tentative bound from the CMS analysis by assuming that \emph{i)} the vertex reconstruction efficiencies $\etrig(v) \times \eoff(v)$ are independent between different vertices and that \emph{ii)} requiring an additional vertex suffices to reduce the background to negligible levels for the whole range of $m_{A'}$. If both assumptions are satisfied, one obtains the yellow curve in Fig.~\ref{fig:ctauplot}. We see that demanding an additional vertex may be a very powerful handle at low to moderate $c\tau_{A'}$.
At higher values of $c\tau_{A'}$, a larger fraction of the long-lived $A'$ escape the tracker before decaying, so that experimental sensitivity is dominated by the single-dimuon vertex search.

\begin{figure*}[t]
\includegraphics[width=0.9\textwidth]{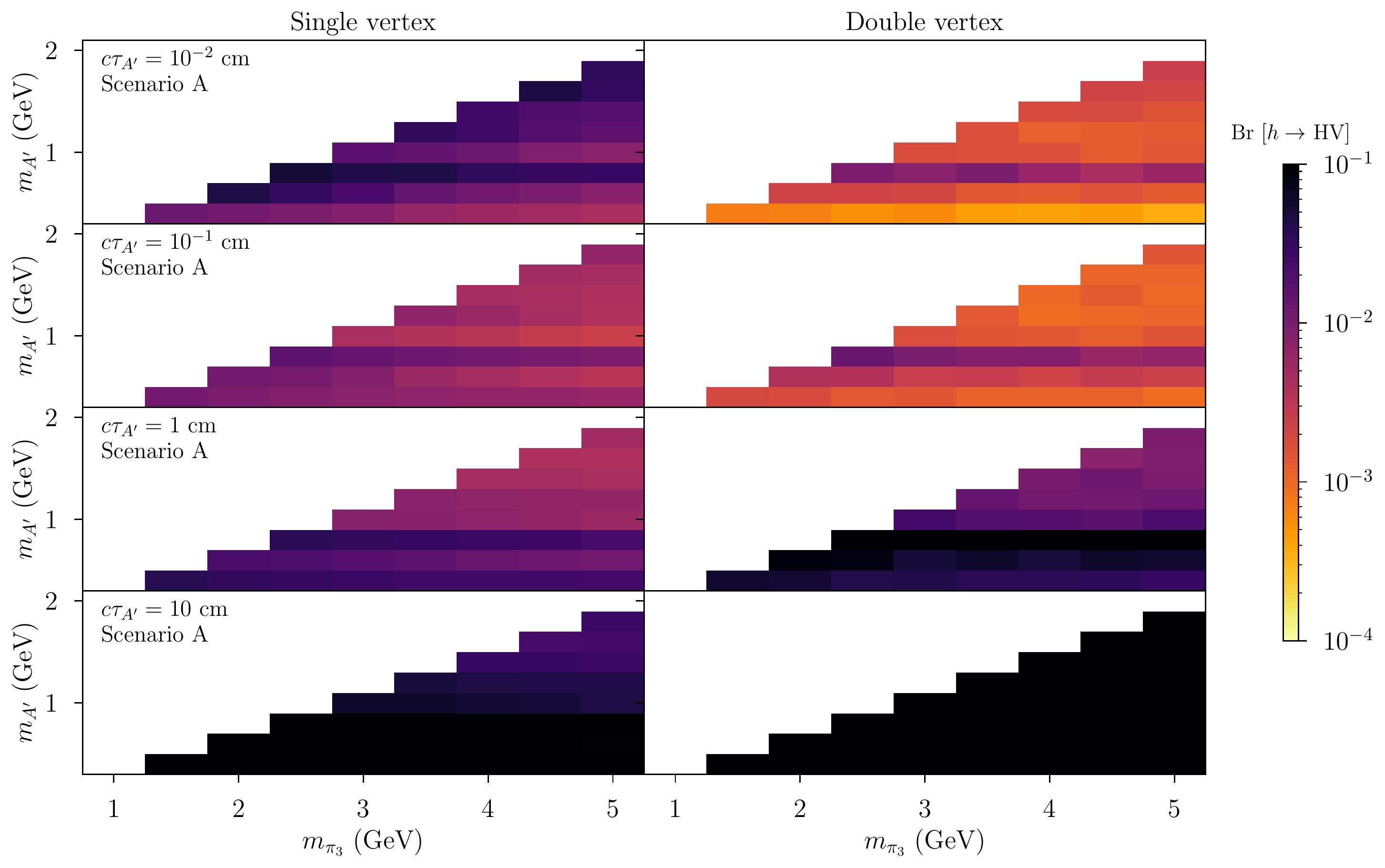}
\caption{Limits on the Higgs branching ratio  to the hidden valley model in Scenario A
from our recast of Ref.~\cite{CMS:2021sch}, with $\Lambda=m_{\eta}=4 m_{\pi_3}$, $\sin\theta=0.1$ and $g=0.05$. (See Appendix~\ref{app:mesonsector} for details.)  The ``single vertex'' column indicates the single vertex isolated or inclusive signal region, whichever gives the strongest limit for the point in question. The ``double vertex'' limit makes important and unvalidated assumptions about correlations in the vertex reconstruction efficiencies and the SM background, as described in the text. It should therefore be interpreted as a sensitivity estimate rather than as a robust limit. 
\label{fig:triangle}}
\end{figure*}

Fig.~\ref{fig:triangle} shows how the sensitivity of the CMS scouting search changes as we vary both $m_{ \pi_3}$ and $m_{A'}$ while keeping $m_{\pi_3}/\Lambda$ fixed. Varying $m_{\pi_3}$ has two effects: a larger $m_{ \pi_3}$ reduces the available phase space for dark mesons, yielding a lower vertex multiplicity. The larger mass however provides a somewhat higher boost to the $A'$ and therefore the muons, increasing the likelihood that said vertices will be reconstructed. Varying $m_{A'}$ primarily affects the $A'\to \mu\mu$ branching ratio, when it comes to signal acceptance;  however, $m_{A'}$ also plays a primary role in determining the background, which is larger at lower dimuon invariant masses.
From Fig.~\ref{fig:triangle} we conclude that the dependence on $m_{\pi_3}$ is fairly mild, while the results are more sensitive to $m_{A'}$. In particular, we see a dip in sensitivity when $m_{A'}$ is close to the mass of the SM $\rho$ meson, since resonant mixing with the SM $\rho$ enhances the \mbox{$A'\to\pi^+\pi^-$} partial width  and accordingly suppresses the branching ratio of the dark photon into muons.  In this figure we again see that a double vertex analysis is particularly powerful at low $c\tau_{A'}$ and loses sensitivity faster at high $c\tau_{A'}$, as expected.

While we have studied a specific benchmark dark shower model,  thanks to the inclusive nature of both the LHCb and CMS searches, we can draw several broad conclusions. 
In particular, we see that the scouting search provides the leading sensitivity to low-mass dark shower signatures when the muons are produced in the decay of a parent particle with lifetime \mbox{$c\tau \gtrsim $ mm}, and for \mbox{$c\tau \gtrsim$} cm the sensitivity is likely dominated by final states with a single dimuon vertex.  Since the scouting search depends only on the kinematics and multiplicity of dimuon pairs, these properties will hold broadly across a range of dark shower models that produce low-mass, muon-philic final states, including those considered in \cite{Pierce:2017taw,Cheng:2019yai, Cheng:2021kjg}. 
The relative advantage of single- vs double-vertex searches will depend weakly on the mass of the mediator initiating the shower as well as the dark gauge coupling, both of which impact the multiplicity of hidden hadrons. 
However, acceptance alone ensures that in the longer-lifetime regime the single-vertex search will come to dominate the sensitivity. We can thus conclude that scouting searches are a uniquely sensitive tool for discovering low-mass hidden valley models, providing leading sensitivity in the well-motivated but not universal cases where such models give rise to displaced dimuon resonances  that point back to the primary vertex. A double-vertex search would moreover provide important additional sensitivity at intermediate parent lifetimes.

\subsection{Scenario B\label{sec:scenarioB}}

Both CMS and LHCb have selection criteria that require the vector sum of the three-momenta of the muons in a given vertex to point back to the beam line, as this is a good handle to reduce background from fake vertices and material interactions. 
However, this pointing cut has poor efficiency on signals where a long-lived state decays to a dimuon pair plus one or more other states. Scenario B presents two such example cases, where we assume that the $A'$ decays promptly, but that the dark meson $\pi_3$ (scenario B1) or $\eta$ (scenario B2) has a macroscopic lifetime. 

\begin{figure}
\includegraphics[width=0.48\textwidth]{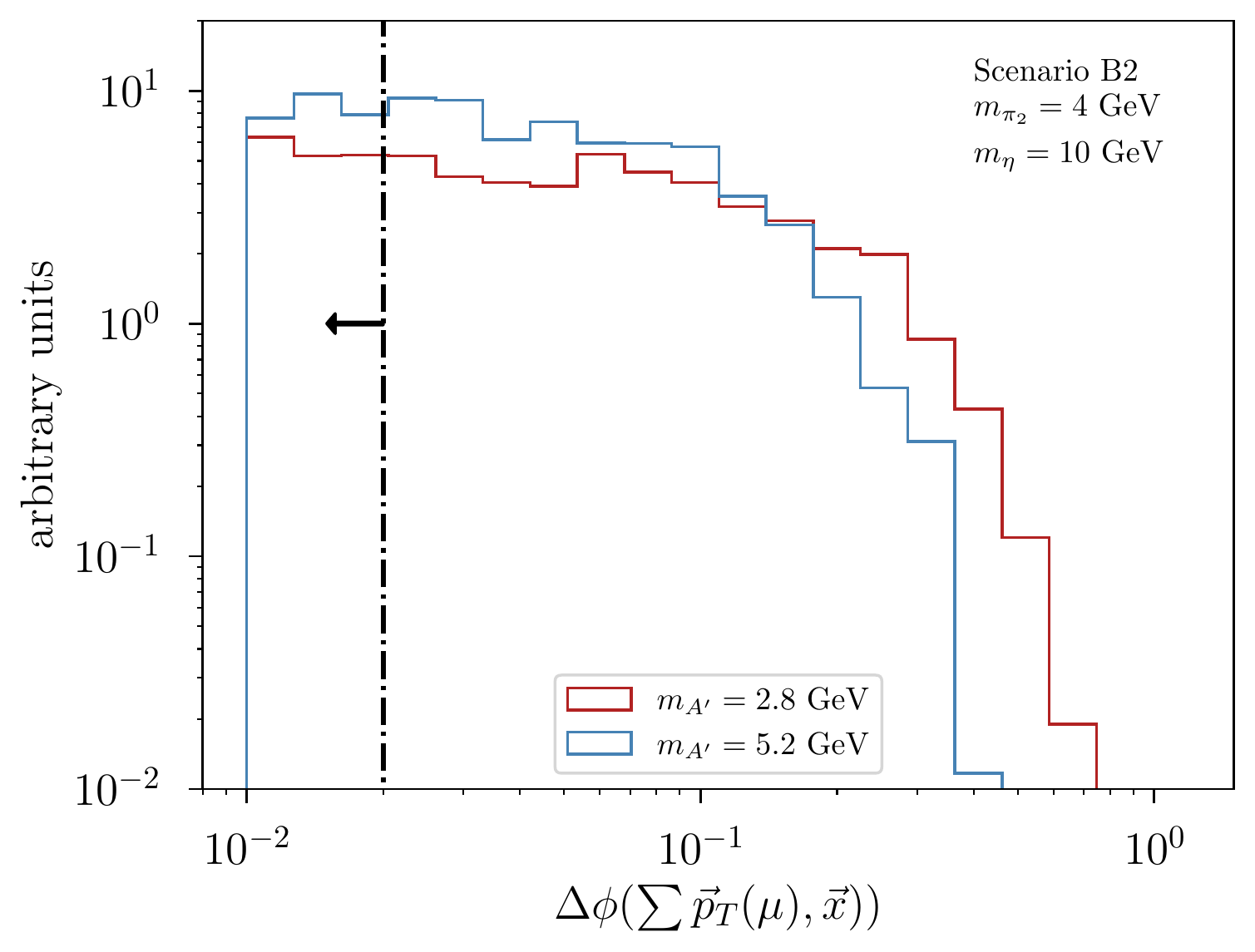}
\caption{ Truth-level distribution of $\Delta \phi(\sum \vec p_T(\mu),\vec x)$ for two example benchmark points in scenario B2. The dashed line indicates the cut imposed in the CMS scouting analysis \cite{CMS:2021sch}. The efficiency of this cut for the red (blue) benchmark is 15\% (10\%). 
\label{fig:scenarioB}}
\end{figure}

We can characterize the degree of pointing through the variable $\Delta \phi(\sum \vec p_T(\mu),\vec x)$, which at truth-level corresponds to the the azimuthal angle $\Delta \phi$ between the three-momenta of the $A'$ and its parent meson,
as shown in Fig.~\ref{fig:scenarioB} for two example points in scenario B2. Though the distribution of $\Delta \phi(\sum \vec p_T(\mu),\vec x))$ is still peaked at small values, 
it is rather broad and the CMS analysis cut (dot-dashed line) is only $\mathcal{O}(10\%)$ efficient. LHCb on the other hand also reports results for an inclusive selection that does not require pointing. This selection comes roughly with an order of magnitude more background than the selection requiring pointing, which is acceptable if a large enough increase in signal efficiency can be achieved.

\begin{figure}
\includegraphics[width=0.48\textwidth]{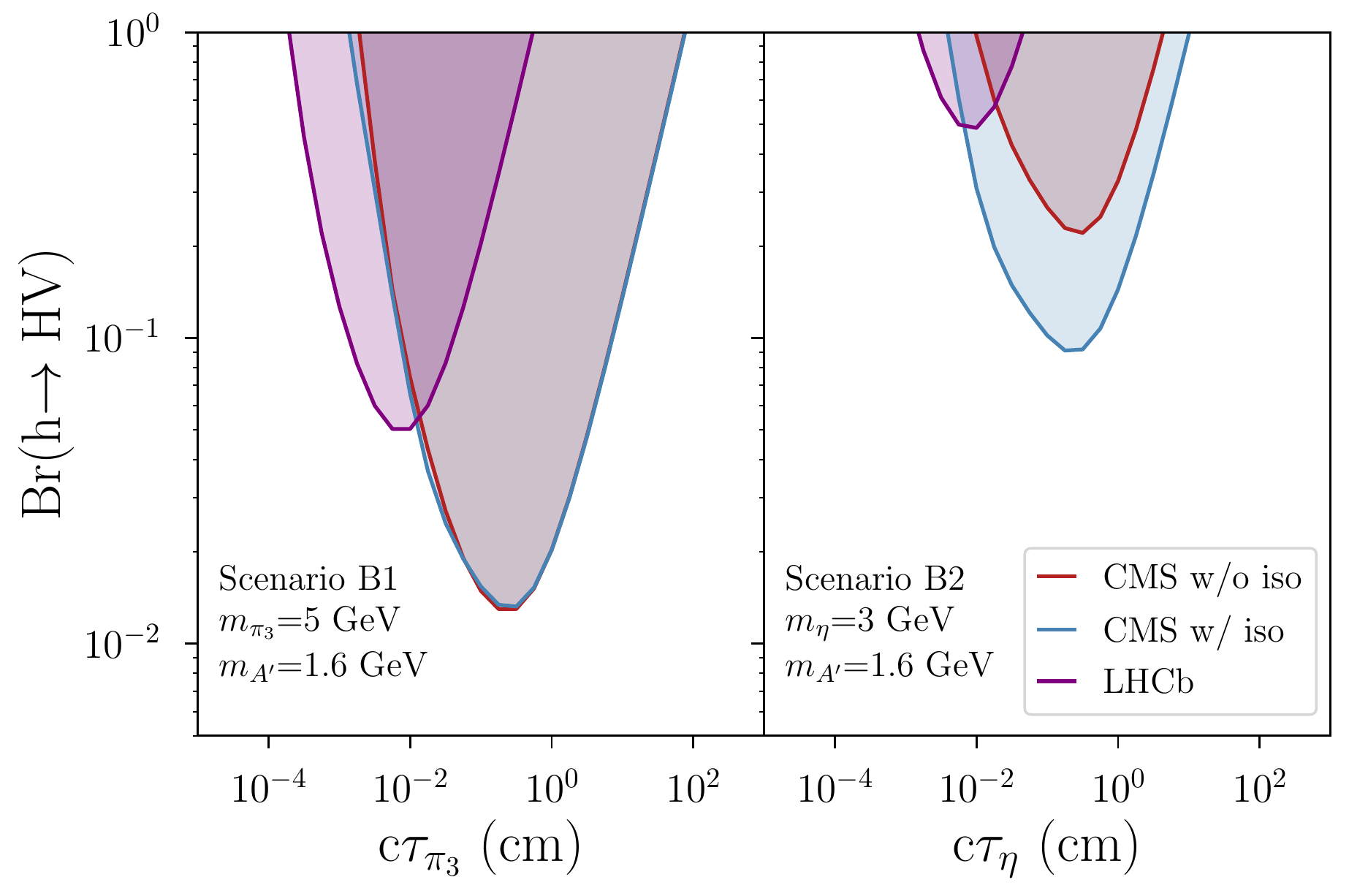}
\caption{ Limits on the branching ratio for the SM Higgs decaying to the hidden valley models described in Sec.~\ref{sec:model}, with $\Lambda=m_{\eta}=4 m_{\pi_3}$, $\sin\theta=0.1$ and $g=0.05$. (See Appendix~\ref{app:mesonsector} for details.) The bounds are shown as a function of the lifetime of the dark photon $c\tau_{A'}$, from our recast of the CMS dimuon scouting analysis \cite{CMS:2021sch} and the LHCb low mass dimuon analysis \cite{Aaij:2020ikh}.}
\label{fig:scenarioBmoney}
\end{figure}

We show the resulting bounds on both Scenarios B1 and B2 in Fig.~\ref{fig:scenarioBmoney} for two example mass points.
As before, blue and red shaded regions correspond to the bounds set by the CMS analysis with and without imposing isolation, respectively. The isolation efficiency for Scenario B1 is comparable to that for Scenario A, roughly 40\%, while for Scenario B2 it is nearly 100\%. 

Since the multiplicity and kinematics of dark photons produced in Scenarios A and B1 are very similar, the differences between the bounds on these two scenarios are mainly attributable to the dimuon pair not pointing to beamline.
Concretely, the LHCb limit on Scenario B1 using the inclusive selection is generally weaker than the Scenario A limit, which uses the pointing selection at the same dark photon mass, owing to larger backgrounds.  The relative stringency of the limits based on these two selections varies bin-by-bin, reflecting background fluctuations; for the specific mass point shown in Fig.~\ref{fig:scenarioBmoney}, the Scenario B1 and Scenario A limits are comparable.
Meanwhile for the CMS scouting search, the loss of acceptance from the cut on $\Delta \phi(\sum \vec p_T(\mu),\vec x)$ weakens the Scenario B1 limit by roughly a factor of five compared to Scenario A.  

For Scenario B2, the number of dimuon pairs per event is substantially smaller, and accordingly the sensitivity of the existing analyses is further reduced. In the bulk of parameter space for this scenario, the existing CMS and LHCb searches do not have better sensitivity than that offered by indirect constraints from global fits to Higgs properties, which currently constrain the exotic Higgs branching fraction to be less than 15\% \cite{ATLAS:2021vrm}.
It would be very interesting for the CMS collaboration to attempt a search with relaxed pointing criteria, as the enhanced signal efficiency may be sufficient to offset the higher backgrounds, particularly for isolated dimuon pairs.

\subsection{Scenario C}

\begin{figure}
\includegraphics[width=0.48\textwidth]{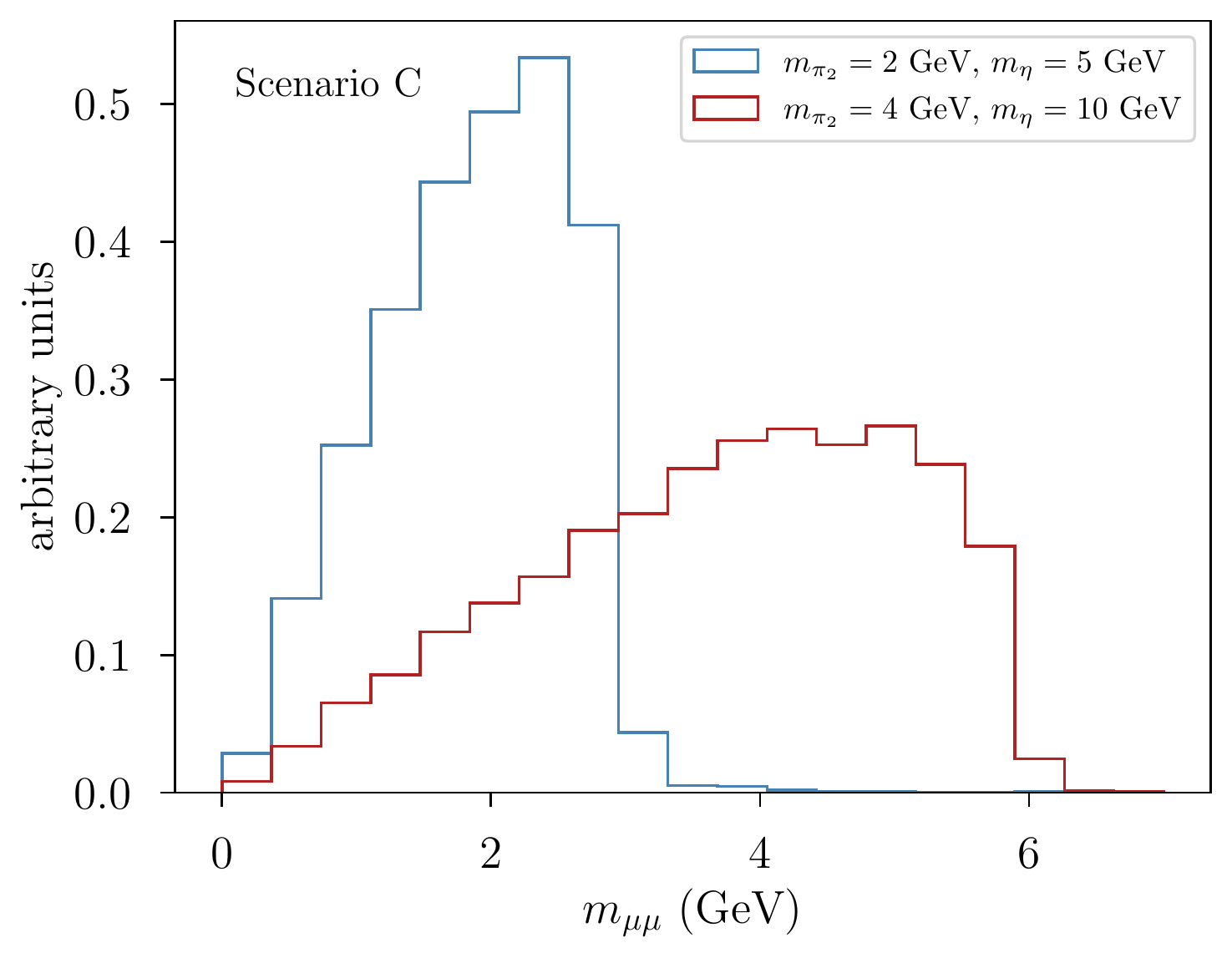}
\caption{Truth-level invariant mass of dimuon pair in scenario C, for two benchmark model points with cm-scale proper decays lengths.  \label{fig:scenarioC}}
\end{figure}

Finally, in scenario C, long-lived $\eta$ mesons produce a displaced dimuon pair that is {\em non-resonant} and hence necessarily non-pointing.
The kinematics of the decay produce an endpoint in the dimuon mass at $m_{\mu\mu,\mathrm{max}} = m_{\eta}-m_{\pi_2}$, as shown in Fig.~\ref{fig:scenarioC}.
Because of the off-shell dark photon and the three-body final state, the minimum possible $\eta$ lifetime for this case is a strong function of the mass splitting between the $\eta$ and the $\pi_2$. For splittings of the order $m_{\eta}-m_{\pi_2} \sim$ few GeV, the minimum possible proper decay length is $\mathcal{O}$(cm), where the exact value depends on the remaining model parameters (see Appendix~\ref{app:branchingratios}). Very narrow splittings ($m_{\eta}-m_{\pi_2} \ll$ GeV) are therefore likely to have intractably low acceptance. 

As for scenario B, we are unable to reliably model the backgrounds for CMS in the absence of the pointing cut and therefore do not attempt to make a sensitivity projection. 
Of course, it is clear this is a much more challenging signal than Scenario B, as the distribution of the signal events over a range of dimuon masses both dilutes the overall statistical significance of the signal as well as complicates the data-driven background estimation process.  An interesting open question at this juncture is to compare the sensitivities of analyses based on scouting triggers to those using alternative trigger strategies, which may have access to a smaller number of signal events, but may contain more information that can be used to control backgrounds.\footnote{Alternatively, one may attempt to expand the amount of useful information recording in the scouting analysis, e.g.~by making use of machine learning-driven data compression methods \cite{Collins:2022qpr}.} The parked data set \cite{CMS-DP-2019-043} is a particularly intriguing possibility in this context. For the more traditional triggers, the relative sensitivities will necessarily depend in detail on the production mechanism. 

As a concrete example, our benchmark scenario of production in exotic Higgs decays  offers the chance to trigger on Higgs production in association with a leptonically-decaying $W$ or $Z$ boson using prompt lepton triggers.\footnote{The associated $ttH$ production cross-section, which is $0.5$ pb at $\sqrt{s} = 13$ TeV \cite{LHCHiggsCrossSectionWorkingGroup:2016ypw}, can provide a sizeable contribution to Higgs production in association with one or more prompt leptons. This production channel is often a challenge for SM analyses owing to the combinatoric challenges of its final states, but it can be important for recording beyond-the-SM Higgs decays to exotic final states.}  While the Higgs production cross-section in semi-leptonic processes is two orders of magnitude smaller than the inclusive Higgs production cross-section relevant for scouting, events that arrive on a standard lepton trigger will contain  information about the whole event and will be subject to different backgrounds.  Without being able to reliably model the backgrounds for non-pointing dimuon pairs, we cannot determine which analysis strategy offers the best sensitivity.

We can, however, conclude that the Higgs production channel considered here is perhaps the {\em most} optimistic choice when it comes to alternative trigger pathways for low-mass dark showers.  Other natural choices for light ($\lesssim$ few hundred GeV) mediators are \emph{i)} the $Z$ boson \cite{Cheng:2019yai,Cheng:2021kjg} and \emph{ ii)} a new SM-singlet, whether vector or scalar (e.g.~\cite{Pierce:2017taw}); one could straightforwardly extend our benchmark model to cover either case by adding additional heavy states, without substantially changing the dark meson phenomenology.  For a $Z$ boson mediator, associated $WZ$ production is more suppressed compared to inclusive $Z$ production than $VH$ is relative to inclusive $H$ production; meanwhile for a BSM mediator such as a $Z'$ or a new scalar, the best additional handle that one could generically expect is energetic initial state radiation (ISR). In this case, the presence of a moderately-hard ISR jet could enable a MET or a standard dimuon trigger strategy, depending on the specifics of the model, at the cost of substantially (and perhaps prohibitively) reducing the available signal cross-section.  An example of this strategy was very recently deployed in the context of a search for inelastic dark matter \cite{CMS-PAS-EXO-20-010}, which relied on a MET trigger and additionally required a soft displaced dimuon pair in the final state.  As this search indicates, the thresholds necessary for a MET trigger strategy are not small; Ref.~\cite{CMS-PAS-EXO-20-010} placed an offline analysis cut of $\met>200$ GeV.
  
\section{Summary and recommendations}
\label{sec:conclusions}

The search for hidden valleys at the LHC is a challenging and multi-faceted program, thanks in part to the fantastic diversity of possible signatures.  Fortunately, search strategies for models with a dark confinement scale at the few GeV scale admit a certain amount of streamlining: many of the portal operators that could govern decays of GeV-scale dark hadrons to SM final states predict lifetimes that are too long to leave much visible energy within the main detectors \cite{Knapen:2021eip}. For such low-mass dark shower events, the operators that are easily compatible with high-multiplicity visible final states tend to be {\em muonphilic}. Thus low-$p_T$ (displaced) dimuon pairs stand out as one of the most promising signatures of low-mass hidden valley theories.

Online analysis techniques, including CMS' data scouting stream and the LHCb low mass dimuon searches, are a powerful tools for probing the soft dimuons characteristic of low-mass dark shower events.  We establish the  sensitivity of CMS' scouting search for displaced dimuon resonances \cite{CMS:2021sch} to dark showers produced in exotic decays of the SM Higgs boson, and compare its reach to that of an inclusive dimuon resonance search at LHCb \cite{Aaij:2020ikh}. We demonstrate the significant advancement in reach realized in the scouting search in the context of a benchmark hidden valley model; however, the inclusive nature of the experimental searches lets us make several observations that hold across a broader class of low-mass dark shower signatures.  Concretely, CMS performs best for medium to long proper lifetimes due to its larger integrated luminosity, while LHCb currently sets the strongest bounds at short lifetimes.  A reinterpretation of the prompt scouting analysis by CMS \cite{CMS-PAS-EXO-21-005} in terms of hidden valley models would be well-motivated. Though we cannot make sharp statements without more information on the background, we expect that double vertex analysis selection by CMS and potentially by LHCb could further improve the bounds.

Our simple and flexible benchmark model features two flavors of light dark quarks and an elementary dark photon, and is capable of realizing several different dimuon vertex topologies with relatively few parameters.  The simplest scenario occurs when dimuon resonances are produced from the decay of a promptly-produced particle, and thus the reconstructed dimuon momentum points back to the beamline. 
Our model also realizes scenarios where the dimuon pair does not point back to the beamline, as well as a non-resonant displaced dimuon signature from a three-body decay.

For the non-pointing case,  the LHCb search currently has a suitable selection, which has only marginal sensitivity with the current dataset. It would therefore be important for both LHCb and CMS to (continue to) include a non-pointing signal region in future searches with more integrated luminosity. Finally, the non-resonant, non-pointing case is also well-motivated and always implies a macroscopic lifetime. We recommend that it too be included as a possibility in future searches.   It remains an interesting open question whether scouting searches or searches for (e.g.) MET in addition to non-pointing dimuon vertices in events that arrive on higher-threshold trigger streams can provide the best probe of these non-resonant scenarios. However here scouting has the particular advantage that searches are {\em independent} of assumptions about the production mode for these showers; alternate trigger strategies depend more strongly on the model-dependent features of the mediator responsible for initiating the shower.

Online analyses such as scouting therefore offer an unmatched discovery tool for low-mass dark showers, which can otherwise easily evade traditional detection strategies.  Of course, in the event of a compelling excess, recording more features of the events would be a top priority.  The relatively high-multiplicity  final states that can be realized in our benchmark models can also provide other handles to further identify and characterize the signature, in particular displaced hadronic and/or electronic vertices and a moderate amount of missing energy.  Displaced hadronic final states in particular are a locus of discussion for future trigger capabilities at LHCb as well as ATLAS and CMS \cite{Alimena:2021mdu}, and could offer further windows onto this challenging class of signatures.

\acknowledgments
We gratefully acknowledge useful conversations with Hsin-Chia Cheng, Matthew Citron, Zeynep Demiragli, Lingfeng Li, Zoltan Ligeti, Steven Lowette, Mario Masciovecchio, Titus Momb\"acher, Dean Robinson, Ennio Salvioni, Christiane Scherb, Matthew Strassler, Indara Suarez, Mahiko Suzuki, Xabier Cid Vidal, and Michael Williams. We thank Matthew Citron, Lingfeng Li, and Ennio Salvioni for comments on the manuscript. SK would also like to thank Diego Redigolo, Alberto Mariotti and Sam Junius for collaboration on related work. The work of SB was supported by  a grant from the United States-Israel Binational Science Foundation (BSF). The work of SK was supported
by the U.S. Department of Energy, Office of Science under contract DE-AC02-05CH11231. Part of this work was performed at the Aspen Center for Physics, which is supported by National Science Foundation grant PHY-1607611.  JS thanks the MIT Center for Theoretical Physics and Laboratory of Nuclear Science for their generous hospitality during the completion of this work.

\bibliography{dark_showers.bib}

\appendix
\section{A two-flavor hidden valley\label{app:appendix}}

Here we construct the benchmark two-flavor model with a light dark photon used in this work. We first discuss the UV Lagrangian in terms of elementary dark sector quarks and dark gluons, then consider the resulting meson spectrum and use chiral perturbation theory to calculate the relevant branching ratios and decay widths. We end with a discussion of the Pythia implementation.

\subsection{Quark sector\label{app:quarksector}}

We consider two dark sector quarks ($q_i$) and their corresponding anti-quarks ($\bar q_i$), which are respectively in the fundamental and the anti-fundamental representation of the dark sector $SU(N_c)$ strong force. We furthermore introduce a single, weakly coupled $U(1)$ gauge field $A'$. 
We will be interested in the case when this gauge field has a mass.  For scenarios A and B1, where the anomalous decay $\pi_3\to A' A'$ is responsible for the visible signal, one can take this mass to come from a St\"uckelberg mechanism, which would be compatible with extending the model into a ``neutral naturalness'' solution to the hierarchy problem.  In order to realize the distinct vertex topologies in scenarios B2 and C, we additionally include a dark Higgs scalar $\phi$, whose vacuum expectation value will contribute to the mass of $A'$ as well as those of the dark quarks.
The charge assignments are as follows:
\begin{equation}
\begin{array}{c|ccccc}\label{eq:chargetable}
& q_1 & q_2 & \bar q_1 & \bar q_2 & \phi\\[2pt]\hline\\[-8pt]
SU(N_c)& \square & \square  & \overline{\strut \square}  & \overline{\strut \square} & 1\\[2pt]
U(1)& 1&-1 &0& 0&1 
\end{array}
\end{equation}
which, at the renormalizeable level, allow for the following set of interactions 
\begin{align}
\mathcal{L}\supset& i \sum_i q^\dagger_i \slashed{D} q_i + i \sum_i \bar q^\dagger_i \slashed{D} \bar q_i + |D^\mu \phi|^2\label{eq:uvkinetic}  \\
\label{eq:yukawaterms} 
&-\left(\!\begin{array}{c}q_1\\q_2\end{array}\!\right)^T
\left(\!\begin{array}{cc} y_{11}\phi^\dagger&y_{12}\phi^\dagger\\
y_{21}\phi&y_{22}\phi\\
\end{array}\!\right)
\left(\!\begin{array}{c}\bar q_1\\\bar q_2\end{array}\!\right)+ H.c.\\
&- \epsilon e A'_\mu J_{EM}^\mu
\label{eq:EMcurrent}
\end{align} 
with $D_\mu \equiv \partial_\mu + i g Q A'_\mu + i g_s T^a G^a_\mu$. Here $g$ and $g_s$ are respectively the $U(1)$ and $SU(N_c)$ gauge couplings, with $Q$ and $T^a$ the $U(1)$ charge and the $SU(N_c)$ generators respectively. The $G^a_\mu$ are the $SU(N_c)$ gluons. For the mass range of interest here, the dark photon can to good approximation be taken to have a small coupling $\epsilon e$ to the SM electromagnetic current $J_{EM}^\mu$, through the mixing of the SM photon with the $A'$. We further assume that the scalar field develops a vacuum expectation value $v$, such that we can decompose it as
\begin{equation}
\phi = \frac{1}{\sqrt{2}}(v+\varphi)e^{i\tilde a/v}.
\end{equation}
The vacuum expectation value contributes both to the mass of the $A'$ and the masses of the quarks through the interactions in \eqref{eq:uvkinetic} and \eqref{eq:yukawaterms}. Concretely, expanding the kinetic term for $\phi$ we find 
\begin{equation}\label{eq:higgsgaugefield}
 |D^\mu \phi|^2 \supset \frac{1}{2}g^2 v^2 A'_\mu A'^\mu+ g v A'_\mu \partial^\mu \tilde a \left(1+\frac{\varphi}{v}\right)^2
\end{equation}
A Higgs portal coupling between SM and dark Higgses is allowed by all symmetries and provides a natural UV completion of the coupling $h q_i \bar q_i$ responsible for initiating the dark showers, as we now sketch. Introducing a Higgs portal mixing between dark and SM Higgs of the form $\mathcal{L}_{int} = \kappa |\phi|^2 |H|^2$ gives rise to a dark Higgs-SM Higgs mixing angle that in the limit $\kappa \ll 1$ can be expressed in terms of the dark Higgs vev $v$ and mass $m_\varphi$ as 
\begin{equation}
    \theta \approx \frac{\kappa v v_h}{m_\varphi^2 - m_h^2},
\end{equation}
where $v_h$ is the SM Higgs vev.
As a result of this mixing, the SM Higgs $h$ picks up a coupling to dark quarks given by
\begin{equation}
\mathcal{L}_{int} =  \frac{y_{ij}}{\sqrt{2}}\theta h \bar q_{Di} q_{Dj}.
\end{equation}
 We will assume for simplicity that the radial mode $\varphi$ decouples from the subsequent phenomenology, and neglect it hereafter.\footnote{Adding the Higgs portal coupling $\kappa$ also shifts the dark quark masses by a quantity of order $\kappa v_h^2/m_\varphi^2$.  For our purposes we can absorb this shift into a redefinition of the quark mass parameters.} In the absence of chiral symmetry breaking, we could simply remove the Goldstone boson $\tilde a$ by working in the unitary gauge. The mass of $A'$ however receives a contribution both from $v$ and from the confining $SU(N_c)$ dynamics, which means that a linear combination of the elementary Goldstone mode $ \tilde a$ with one of the meson modes will furnish the longitudinal component of the $A'$ in the unitary gauge. This is best treated in chiral perturbation theory, with the meson degrees of freedom; we describe it in the next section.

The quark sector in \eqref{eq:yukawaterms} a priori leaves us with a lot of freedom and is therefore rather unwieldy. Rather than mapping out the phenomenology of the fully general case, we are  interested in picking an example that is simple to parametrize and that generates the signatures we are interested in. For this reason we will unapologetically assume the following relations between the Yukawa couplings in \eqref{eq:yukawaterms}:
\begin{align}
y_{11}=y_{22} \quad \mathrm{and}\quad y_{12}=y_{21}\label{eq:paramchoices}.
\end{align}
To track the gauge invariance in the low-energy chiral Lagrangian, it will be useful to retain the explicit dependence on the goldstone mode $\tilde a$ coming from the fundamental Higgs boson. Thus we write the mass matrix as
\begin{equation}
  M = \Phi M_0,
\label{eq:massmatrixwithphase}
\end{equation}
where  $M_0$ is the mass matrix with $\tilde a=0$, and  $\Phi \equiv \mathrm{diag}(e^{-i \tilde a/v},e^{i \tilde a/v}) $ in the gauge basis.
If we move to the basis that diagonalizes $M_0$, the full mass matrix $M$ then reads 
\begin{equation}
\mathcal{L}\supset -\left(\!\!\begin{array}{c}\bar q_1 \\ \bar q_2\end{array}\!\!\right)^T
\left(\!\!\begin{array}{cc}
 m_1 \cos\left(\frac{\tilde a}{v}\right)&  -i\,  m_2\sin\left(\frac{\tilde a}{v}\right)\\
 -i\,  m_1\sin\left(\frac{\tilde a}{v}\right)&m_2\cos\left(\frac{\tilde a}{v}\right)\\
 \end{array}\!\!\right)
\left(\!\!\begin{array}{c} q_1 \\ q_2\end{array}\!\!\right)+ \mathrm{H.c.}
\label{eq:quarkmassmatrix}
\end{equation}
with $m_1\equiv \frac{v}{\sqrt{2}}(y_{11}-y_{12})$  and  $m_2\equiv \frac{v}{\sqrt{2}}(y_{12}+y_{11})$. We explicitly retain the phase $\tilde a/v$, as it will play a role in the chiral perturbation theory calculations in the next section.

At this point it is convenient to switch to Dirac notation, which will make the approximate $SU(2)\times SU(2)$ flavor symmetry manifest and facilitate the matching onto chiral perturbation theory. The model is then specified by (with $\tilde a=0)$
\begin{equation}
\mathcal{L}\supset i\sum_{i=1,2} \overline Q_i \slashed{D} Q_i - m_i \overline Q_iQ_i
\end{equation}
with the Dirac fermions $Q_i=(q_i \; \bar q_i^\dagger)^T$. In this basis, the interactions of the $A'$ are now manifestly chiral, and the covariant derivative is defined as 
\begin{equation}
D_\mu \equiv \partial_\mu + i g Q_L P_L A'_\mu + i g Q_R P_R A'_\mu + i g_s T^a G^a_\mu
\end{equation}
with the $P_{L,R}$ the chiral projection operators. The charge matrices $Q_{L,R}$ are obtained by rotating the charges in \eqref{eq:chargetable} to the basis that diagonalizes $M_0$:
\begin{equation}
Q_L=\left(\!\!\begin{array}{cc}
&1\\
1&\\
 \end{array}\!\!\right)\quad \mathrm{and} \quad Q_R=0.
 \label{eq:quarkchargematrix}
\end{equation}
In general, the charge matrices in the $M_0$ basis can be non-sparse and contain non-integer numbers, as does the CKM matrix in the SM. In other words, the simple form of \eqref{eq:quarkchargematrix} is non-generic and a consequence of our choices in \eqref{eq:paramchoices}. This choice is intended to streamline the analysis in the next section, without qualitatively changing the phenomenology. Note that the $A'$ interaction breaks both $C$ and $P$ symmetry, which is essential to realize the $\eta \to A' \pi_2$ decay channel. It preserves $CP$, however, and we will assume that the $\theta$-angle associated with the $SU(N_c)$ dynamics is small enough to not meaningfully affect the phenomenology, as is the case in the SM.

\subsection{Meson sector\label{app:mesonsector}}
\newcommand{\bpi}{\boldsymbol\pi}

We assume that gauge coupling and masses in Sec.~\ref{app:quarksector} are a small perturbation on the strong dynamics in the dark sector. In this case, the lightest dark sector mesons are well described as the three pseudo-goldstone bosons associated by the breaking of the \mbox{$U(2)\times U(2)\to U(2)$} symmetry by the dynamics of the confining gauge group. The mesons ($\tilde \pi_i$) therefore make up the adjoint representation of the unbroken $U(2)$, such that we can define the following matrix of meson fields 
\begin{equation}\label{eq:gellmann}
\bpi \equiv \sum_{i=0}^3 \lambda_i \tilde\pi_i
\end{equation}
with the $\lambda_i$ the Pauli matrices for $i=1,2,3$ and $\lambda_0=\mathbb{1}_{2\times2}$. In other words, the $\tilde\pi_i$ refer to the fields in the symmetry eigenbasis, as defined by \eqref{eq:gellmann}. We will reserve the $\pi_i$ notation for the fields in the mass eigenbasis, as introduced below.
We further define the matrix of fields 
\begin{equation}\
\Sigma \equiv \frac{1}{2} f e^{i \bpi/f}.
\end{equation}
This field transforms as $\Sigma \to L^\dagger \Sigma R$ under the $U(2)\times U(2)$ flavor symmetry of the UV theory and as  $\Sigma \to V^\dagger \Sigma V$ under the unbroken $U(2)$. The meson effective theory is then described by
\begin{equation}\label{eq:chilag}
\mathcal{L} = \mathcal{L}_{\mathrm{kin}}+\mathcal{L}_{\mathrm{mass}}+\mathcal{L}_{\mathrm{an}}
\end{equation}
with
\begin{align}
\mathcal{L}_{\mathrm{kin}}&= \mathrm{Tr}\left[D^\mu \Sigma^\dagger D_\mu \Sigma\right] \label{eq:chilagkin} \\
\mathcal{L}_{\mathrm{mass}}&= \ c_m \Lambda f \mathrm{Tr}\left[M \Sigma\right]+ \mathrm{H.c.}\label{eq:chilagmass} \\
\mathcal{L}_{\mathrm{an}}&= \frac{m_0^2}{8}  \mathrm{Tr}\left[\ln \Sigma - \ln \Sigma^\dagger \right]=-\frac{1}{2}m_0^2 \tilde\pi_0^2\label{eq:chilagann}
\end{align}
with 
\begin{equation}
D_\mu\Sigma \equiv \partial_\mu \Sigma+i g A'_\mu (Q_L\Sigma-\Sigma Q_R).
\end{equation}
Here $M$ is the quark mass matrix defined in Eqs.~\ref{eq:massmatrixwithphase}-\ref{eq:quarkmassmatrix}. The parameter $\Lambda$ in \eqref{eq:chilagmass} represents the dark sector's confinement scale, while $c_m$ is an $\mathcal{O}(1)$, dimensionless matching coefficient. We set $c_m=1$ going forward.\footnote{We will always present our results in terms of the physical meson masses; as such the value of $c_m$ will only enter when converting the $m_{\pi_i}$ to the quark masses. The value of $c_m$ therefore does not affect the phenomenology, except when we verify that the Yukawa couplings in \eqref{eq:yukawaterms} are perturbative for the meson spectra of our choice.}
\eqref{eq:chilagann} reflects the fact the isospin singlet $\pi_0$ is not a true Goldstone boson and receives a mass correction due to instanton effects \cite{Witten:1980sp}, as does the $\eta'$ in the SM. The coefficient $m_0^2$ scales as $1/N_c$ in the large $N_c$ limit. We will assume that $m_0\gg m_{1,2}$ in our analysis. Here we have kept only the leading terms in the momentum and $1/N_c$ expansions, an approximation that reproduces the meson spectrum in the SM to within $\sim 20\%$ accuracy (see e.g.~\cite{Degrande:2009ps}). 

We can obtain the mass matrix for the mesons by expanding \eqref{eq:chilagmass} and \eqref{eq:chilagann} to leading non-trivial order in $1/f$ and \eqref{eq:quarkmassmatrix} to second order in $1/v$. This results in
\begin{align}
\mathcal{L}\supset&-
 \frac{1}{2}\bar m \Lambda \left(\!\!\begin{array}{c} \tilde a \\ \tilde\pi_1 \\ \end{array}\!\!\right)^T
 \left(\begin{array}{cc}  
\frac{f^2}{v^2}&\frac{f}{v}\\
\frac{f}{v}&1
 \end{array}\!\!\right) 
 \left(\!\!\begin{array}{c} \tilde a\\ \tilde\pi_1  \end{array}\right)
 -  \frac{1}{2}\bar m \Lambda \tilde \pi_2^2 \nonumber \\
&- \frac{1}{2}\left(\!\!\begin{array}{c} \tilde\pi_0 \\ \tilde\pi_3 \end{array}\!\!\right)^T
 \left(\begin{array}{cc}  
 m_0^2 +\bar m \Lambda & \delta m \Lambda\\
 \delta m \Lambda &\bar m \Lambda 
 \end{array}\!\!\right)
 \left(\!\!\begin{array}{c} \tilde\pi_0 \\ \tilde\pi_3  \end{array}\right)\label{eq:mesonmasses}
\end{align}
with $\bar m \equiv m_1+m_2$ and $\delta m \equiv m_2-m_1$. The isospin-conserving limit is retrieved by sending $\delta m\to 0$ and $v\to \infty$, in which case we find a light, degenerate isospin triplet, analogous to the SM pions, and a heavy isospin singlet, analogous to the SM $\eta$. We will however work with broken isospin, i.e, with $\delta m\neq 0$, such that the $\tilde\pi_0$ and $\tilde\pi_3$ eigenstates mix. We can define the fields in the mass basis as
\begin{align}
a&\equiv \cos\phi\, \tilde a + \sin\phi\, \tilde \pi_1 \\
\pi_1&\equiv \sin\phi\, \tilde a - \cos\phi\, \tilde\pi_1\\
\pi_2&\equiv \tilde \pi_2\\
\pi_{3}& \equiv \cos \theta\, \tilde\pi_3 + \sin \theta\,  \tilde\pi_0 \\
\eta &\equiv \sin \theta\,  \tilde\pi_3 - \cos \theta\,  \tilde\pi_0
\end{align}
with the mixing angles specified by
\begin{align}
\tan \phi &=\frac{f}{v}\\
\tan \theta &= \frac{\sqrt{m_0^4 + 4 \delta m^2 \Lambda^2}-m_0^2}{2 \delta m \Lambda}\\
& \approx \frac{\Lambda \delta m}{m_0^2}+ \mathcal{O}\left(\frac{\delta m^3 \Lambda^3}{m_0^6}\right)
\end{align}
The masses are
\begin{align}
m_{a}^2&= 0\\
m_{\pi_{1}}^2&= \bar m \Lambda\left(1+\frac{f^2}{v^2}\right)\\
m_{\pi_{2}}^2&= \bar m \Lambda\\
m_{\pi_3}^2&= \bar m \Lambda\left(1-\tan \theta \frac{\delta m}{\bar m}\right)\\
m_{\eta}^2&= m_0^2+ \bar m \Lambda\left(1+\tan \theta \frac{\delta m}{\bar m}\right)\end{align}
The massless linear combination $a$ will furnish the longitudinal component of the $A'$ in the unitary gauge, as we show below. The isospin-preserving limit corresponds to $\sin\theta \to 0$ and $v\to \infty$.

Turning now to the kinetic term in \eqref{eq:chilagkin}, we can add the kinetic term for the Higgs field in \eqref{eq:higgsgaugefield}, expand in $1/f$ and subsequently move to the mass eigenbasis. Aside from canonical kinetic terms for the $a$, $\pi_i$ and $\eta$ fields, this yields the following terms
\begin{align}\label{eq:chilagexpanded0}
\mathcal{L}\supset &\frac{1}{2} m_{A'}^2 A'^\mu A'_\mu +  m_{A'} A'^\mu \partial_\mu a \\
\label{eq:chilagexpanded}
&+  g \sin \theta A'^\mu \left(\pi_2 \partial_\mu \eta - \eta \partial_\mu \pi_2  \right)\\
\label{eq:chilagexpanded2}
&+g \cos \theta A'^\mu\left(\pi_3 \partial_\mu \pi_2 - \pi_2 \partial_\mu \pi_3  \right) +\cdots
\end{align}
with $m_{A'}\equiv g \sqrt{v^2+f^2}$. The $\cdots$ represent higher order terms in the $1/v$ and $1/f$ expansions. We see that the $A'$ mass term receives a contribution from the chiral condensate, as expected. The massless linear combination $a$ moreover indeed corresponds to the mode that is eaten by the $A'$ in the unitary gauge.

As long as isospin is broken ($\sin\theta \neq 0$), the \mbox{$\eta \to A' \pi_2$} decay mode is available. This decay is both $C$ and $P$ violating, similar to the \mbox{$\eta' \to \rho \pi_0$} mode in the SM. Unlike the SM, the UV completion we chose in Appendix~\ref{app:quarksector} maximally breaks both $C$ and $P$, such that the \mbox{$\eta \to A' \pi_2$} decay in our dark sector is allowed. The decay will always proceed to the longitudinal component of the $A'$, since the decays to the transverse polarizations are  incompatible with angular momentum conservation.

\subsection{Decay rates and branching ratios\label{app:branchingratios}}
We consider three decay modes for the dark sector mesons $\pi_3$ and $\eta$ that can give rise to displaced muon pairs:
\begin{enumerate}
\item $\pi_3,\eta \to A'A'$ through the $A'$ chiral anomaly.
\item $\eta \to \pi_2 A'$ through \eqref{eq:chilagexpanded}.
\item $\eta \to \pi_2 A'^\ast \to \pi_2 f\bar f$ through \eqref{eq:chilagexpanded} as a three-body decay, with the $A'$ off-shell.
\end{enumerate}
For simplicity, we will always choose our mass benchmark points such that other pion decay modes such as \mbox{$\pi_2\to \pi_3 A'$} are kinematically closed.  The other dark mesons $\pi_{1,2}$ are detector-stable and contribute to missing energy.

The \mbox{$\eta \to 3\pi$} decay is isospin-violating, but otherwise proceeds through the dark sector's strong interaction. Provided it is kinematically allowed and $\sin\theta$ is not tiny, we therefore assume that this decay happens promptly.
The leading terms in the chiral Lagrangian, Eqs.~\ref{eq:chilag}--\ref{eq:chilagann}, predict equal branching ratios to the final states $\pi_1\pi_1 \pi_3$ and $\pi_2\pi_2\pi_3$, up to phase space corrections due to the small mass difference between the pions. However, these branching ratios will be corrected by the higher-order operator $\mathrm{Tr}\left(D_\mu \Sigma^\dag D^\mu \Sigma (M \Sigma + \mathrm{h.c.})\right)$, which contributes at the same order in isospin-breaking and derivatives, and whose coefficient is a priori undetermined.  For definiteness and simplicity, when the $\eta\to 3\pi$ mode is kinematically open, we will assume
\begin{equation}
\text{Br}[\eta\to \pi_1\pi_1 \pi_3]=\text{Br}[\eta\to \pi_2\pi_2 \pi_3]=\frac{1}{2},
\end{equation}
again up to phase space corrections.  

The $\pi_3,\eta \to A'A'$ decays are mediated by the anomalous current
\begin{align}
\partial_\mu J^{0\mu}_5 &= -\frac{\alpha'}{8\pi} N_c \mathrm{Tr}\left[\lambda^0 Q_L^2\right]F'^{\mu\nu}\tilde F'_{\mu\nu}\\
&= -\frac{\alpha'}{4\pi} N_c F'^{\mu\nu}\tilde F'_{\mu\nu}
\end{align}
where in the first line the trace runs only over flavor indices, and $\alpha'\equiv g^2/4\pi$. We can identify the matrix elements by
\begin{align}
\langle 0 | J^{0\mu}_5 | \pi_3\rangle&=-i \sin\theta f_\pi p^\mu e^{-i \mathbf{x}\cdot \mathbf{p}} \\
\langle 0 | J^{0\mu}_5 | \eta \rangle&=-i \cos \theta f_{\eta} p^\mu e^{-i \mathbf{x}\cdot \mathbf{p}}
\end{align}
with $f_\pi$ and $f_\eta$ the $\pi_3$ and $\eta$ decay constants.
Consequently, we find 
\begin{align}
\Gamma_{\pi_3\to A'A'}&=\frac{\alpha'^2}{64\pi^3}N_c^2 \sin^2\theta \frac{m_{\pi_3}^3}{f^2}\\
\Gamma_{\eta\to A'A'}&=\frac{\alpha'^2}{64\pi^3}N_c^2 \cos^2\theta\frac{m_{\eta}^3}{f^2}.
\end{align}
where for definiteness we set $f_\pi\approx f_\eta \approx f$.  For the $\pi_3$ meson, this is the only decay mode, which proceeds with a proper lifetime of 
\begin{align}
c\tau \sim&\; 1.7\times 10^{-7}\; \mathrm{cm}\times \left(\frac{0.01}{\alpha'}\right)^2\times \left(\frac{0.1}{\sin\theta}\right)^2\nonumber\\
&\times \left(\frac{1\,\mathrm{GeV}}{m_{\pi_3}}\right)^3\times \left(\frac{f}{0.2\,\mathrm{GeV}}\right)^2.
\end{align}
In other words, by dialing the gauge coupling $\alpha'$, one can choose the $\pi_3$ to be either prompt or very long-lived on collider-relevant timescales.

The $\eta$ can a priori also decay through $\eta \to \pi_2 A'$, provided that $m_{\pi_2}+m_{A'}<m_{\eta}$. The partial width of this channel is
\begin{equation}
\label{eq:twobodywidth}
\Gamma_{\eta \to \pi_2 A'}= \frac{\alpha' \sin^2\theta\, m_{\eta}^3}{4 m_{A'}^2} \left( (1-x_\pi^2-x_A^2)^2-4 x_\pi^2 x_A^2 \right)^{3/2}
\end{equation}
with $x_\pi\equiv m_{\pi_2}/m_{\eta}$ and $x_A\equiv m_{A'}/m_{\eta}$. The decay width in \eqref{eq:twobodywidth} appears to diverge in the $m_{A'}\to 0$ limit, which is expected from the goldstone equivalence theorem. Provided that there is no extreme phase space suppression ($x_\pi\ll 1$ and $x_A\ll1$) and that the $\eta\to A'A'$ channel is either subleading or kinematically closed, the $\eta$ lifetime is 
\begin{align}
c\tau \sim&\; 3\times 10^{-11}\; \mathrm{cm}\times \left(\frac{0.01}{\alpha'}\right)\times \left(\frac{0.1}{\sin\theta}\right)^2\nonumber\\
&\times \left(\frac{2\,\mathrm{GeV}}{m_{\eta}}\right)^3\times \left(\frac{m_{A'}}{0.5\,\mathrm{GeV}}\right)^2.
\end{align}
Also here we see that the decay can be made either prompt or displaced, depending on the values adopted for $\alpha'$ and $\sin\theta$. 

\begin{figure}
\includegraphics[width=0.5\textwidth]{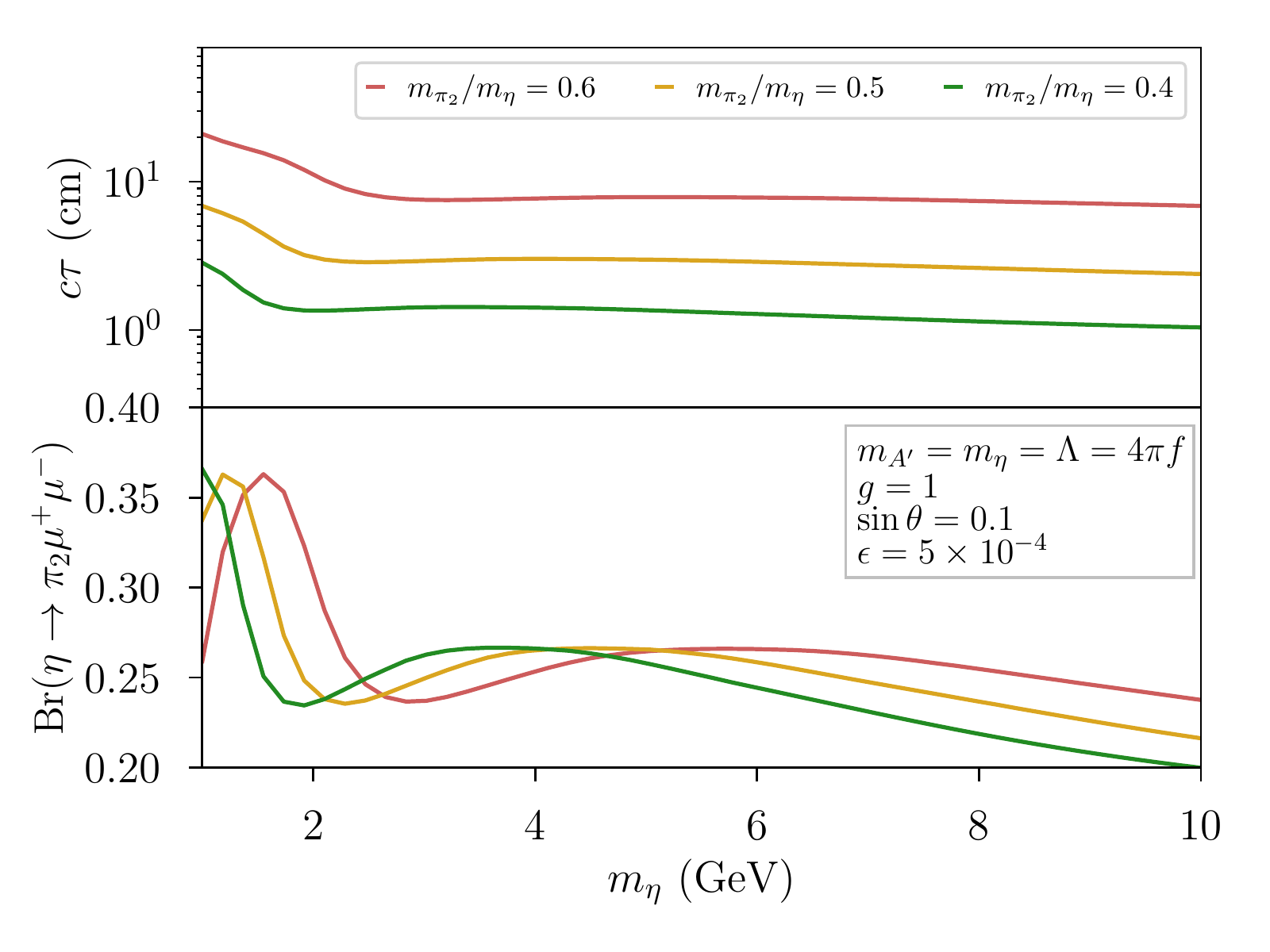}
\caption{Proper decay length and dimuon branching ratio for the three body decay in \eqref{eq:threebodyall}, for a few benchmark model points as indicated in the figure. For $3 m_{\pi_2}<m_{\eta}$, the $\eta\to 3\pi$ decays will always dominate.\label{fig:threebodyplot}}
\end{figure}

Finally, if the $\eta \to \pi_2 A'$ decay is kinematically closed, the three-body decay $\eta\to\pi_2 f\bar f$ will dominate. The partial width to muons for this process is
\begin{align}
\frac{d\Gamma_{\eta\to \pi_2\mu\mu}}{d x} =&\frac{\epsilon^2 \alpha' \alpha  \sin^2\theta }{12\pi}m_{\eta}\frac{x+2x_\mu^2 }{(x-x_A^2)^2}\sqrt{x-4x_\mu^2}\nonumber\\
&\times\left(\frac{(x-1)^2-2(x+1)x_\pi^2+x_\pi^4}{x}\right)^{3/2}\label{eq:threebodymuondiff}
\end{align}
with $\alpha$ the SM fine structure constant, $x_\mu\equiv m_\mu/m_{\eta}$ and $x=q^2/m^2_{\eta}$, where $q^2$ is the invariant mass formed by the muon pair. The partial width and total width are obtained by evaluating
\begin{align}
\Gamma_{\eta\to \pi_2\mu\mu}&=\int_{4 x_\mu^2}^{(1-x_\pi)^2}\!\!\! dx \frac{d\Gamma_{\eta\to  \pi_2\mu\mu}}{d x},\label{eq:threebodymuon}\\
\Gamma_{\eta\to \pi_2 f\bar f}&=\int_{4 x_\mu^2}^{(1-x_\pi)^2}\!\!\! dx \frac{d\Gamma_{\eta\to \pi_2\mu\mu}}{d x} R(x m^2_{\eta})\label{eq:threebodyall}
\end{align} 
with $R(q^2)$ the R-ratio, as extracted from experiment \cite{ParticleDataGroup:2022pth}. Fig.~\ref{fig:threebodyplot} shows the proper decay length and dimuon branching ratio for a set of benchmark points, which were chosen to roughly minimize $c\tau$ without resorting to unphysical choices for underlying couplings. The photon-dark photon mixing parameter $\epsilon$ was chosen to satisfy the existing constraints from direct searches for dark photons~\cite{Gori:2022vri}. We see that this decay tends to be somewhat displaced, with $c\tau\gtrsim 1$ cm. The branching ratio to muons hovers around 25\% and is relatively insensitive to $m_\eta$ and $m_{\pi_2}$. For completeness, we include all underlying parameters for an example point in Tab.~\ref{tab:modelpoint}.

\begin{table}
\begin{tabular}{l|ll}

\multirow{9}{*}{\rotatebox[origin=c]{90}{meson sector}}&$m_{\pi_1}$& 0.403 GeV\\
&$m_{\pi_2}$& 0.4 GeV\\
&$m_{\pi_3}$& 0.396 GeV\\
&$m_{\eta}$& 1 GeV\\
&$m_{A'}$&1 GeV\\
&$\sin\theta$&0.1\\
&$f$&0.080 GeV\\
&$g$&1\\
&$\epsilon$&$5\times 10^{-4}$\\\hline
\multirow{7}{*}{\rotatebox[origin=c]{90}{quark sector}}&$m_1$ & 0.072 GeV\\
&$m_2$ & 0.088 GeV\\
&$\Lambda$ &1 GeV\\
&$m_0$& 1 GeV\\
&$v$&0.997 GeV\\
&$y_{11}$& 0.012\\
&$y_{12}$&0.11\\\hline
& $c\tau$&2.8 cm\\
& $\mathrm{Br}[\eta \to \pi_2 \mu^+\mu^-]$&0.37 
\end{tabular}
\caption{Example model point for which a relatively short-lived $\eta \to \pi_2 \mu^+\mu^-$ decay is realized. \label{tab:modelpoint}}
\end{table}

\subsection{Pythia 8 implementation\label{sec:pythiacard}}

For Monte Carlo implementation of our model, we use the Pythia 8 hidden valley module \cite{Carloni:2010tw,Carloni:2011kk}. Our model relies on  a recent update  \cite{Albouy:2022cin}, specifically the new \verb+separateFlav = on+ flag, which allows us to set the masses of the pions individually. While our work was in the final stages of preparation, Pythia version 8.309 was released, which contains further updates to the hidden valley module. We did not benchmark versions 8.308 and 8.309 against each other, but we did update our python tool to make it compatible with version 8.309 by adding the new required flag \verb+HiddenValley:setLambda = on+. For version 8.308 this line should be commented out in the Pythia configuration cards. In the remainder of this appendix we specify our full settings and mention some approximations and caveats.

Firstly, we note that the dark $\pi^\pm$ with PDG codes $\pm$\verb+4900211+ are each other's anti-particles and therefore degenerate in mass. In our case, the mass eigenstates are the $\pi_1$ and $\pi_2$, which need not be degenerate, as the dark $U(1)$ is broken spontaneously. We therefore always work in the regime where $m_{\pi_1}\approx m_{\pi_2}$ and simply identify those states with the $\pm$\verb+4900211+ states in the Pythia module. 

There is substantial uncertainty in the hadronization probabilities of the dark sector vector mesons, as well as its $\eta$-meson. For the vector mesons, we choose \verb+HiddenValley:probVector=0.75+, such that the hadronization probability matches the naive expectation from the counting of the number of degrees of freedom. For the $\eta$ meson, we fix \verb+HiddenValley:probKeepEta1 = 1.0+, which corresponds to assuming that the suppression from taking $m_{\eta}>m_{\pi}$ in the Lund string model \cite{Andersson:1983ia,Andersson:1983jt} is sufficient to model the $\eta$ hadronization probability.

We set the masses of the vector mesons and the $\eta$ meson to be equal to the confinement scale $\Lambda$ and choose the constituent quark masses (\verb+4900101+ and \verb+4900102+) to be $\Lambda + m_q$, in line with the recommendations in \cite{Albouy:2022cin}. Self-consistent Pythia cards for this model can be generated with our publicly-available python code \cite{code}.

\section{Signal efficiency plots\label{app:appendix_eff}}

We present signal efficiencies for the CMS scouting analysis as a function of the lifetime of the long-lived particle, for four of the example benchmark points discussed in the main text.

\begin{figure*}[p]
    \centering
    \includegraphics[width=\textwidth]{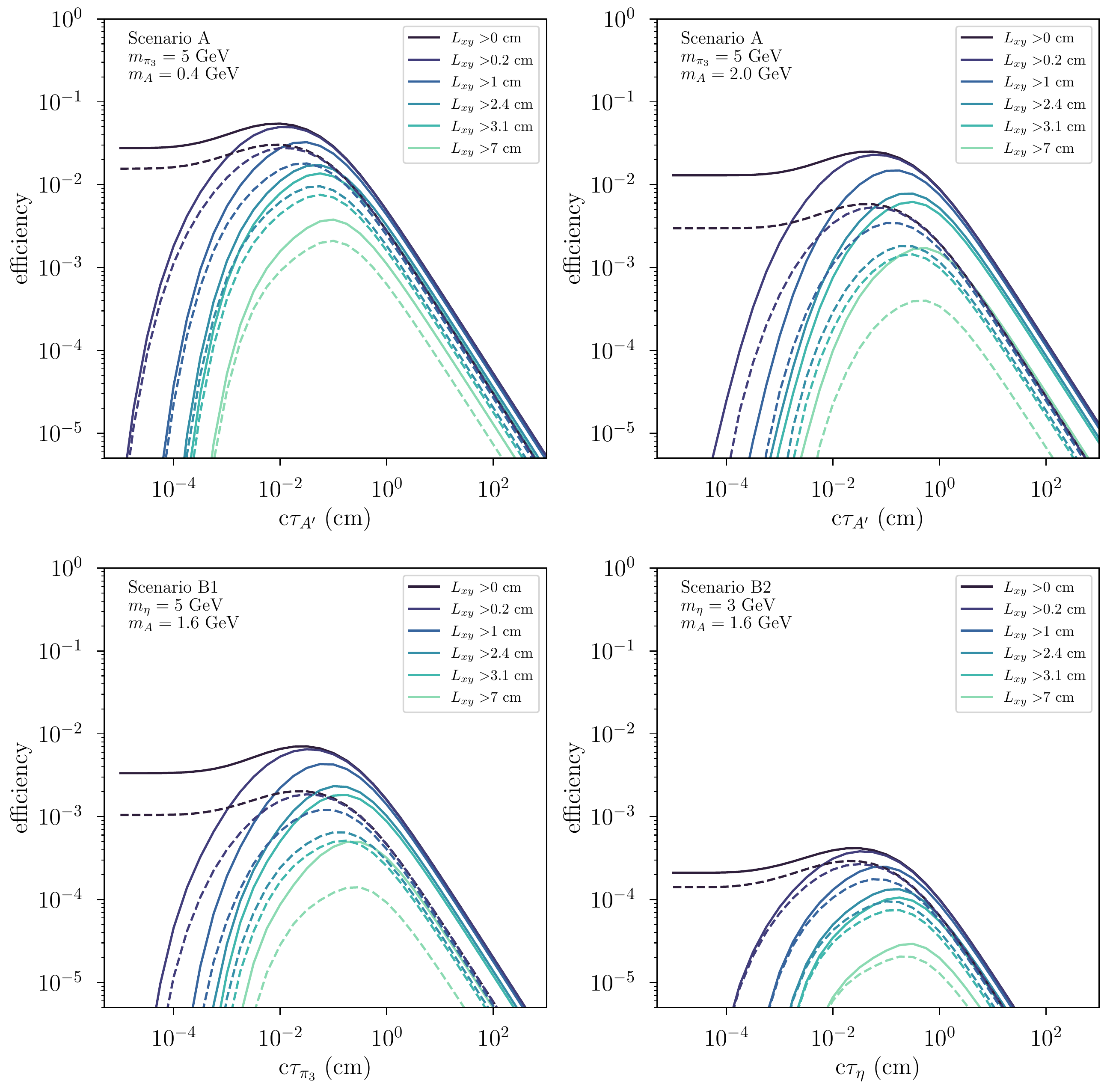}
    \caption{Signal efficiencies for the CMS analysis, for select benchmark points. Solid (dashed) lines are signal are without (with) imposing the isolation condition.}
    \label{fig:eff_appendix}
\end{figure*}

\end{document}